\documentstyle[epsf,aps,pre]{revtex}
\newcommand{\be}{\begin{equation}}
\newcommand{\ee}{\end{equation}}
\newcommand{\bea}{\begin{eqnarray}}
\newcommand{\eea}{\end{eqnarray}}
\makeatother

\begin{document}

\title{Instabilities of Hexagonal Patterns with Broken Chiral Symmetry}

\author{Blas Echebarria and Hermann Riecke\\
 {\small Department of Engineering Sciences and Applied Mathematics,}\\
 {\small Northwestern University, 2145 Sheridan Rd}\\
 {\small Evanston, IL, 60208, USA}\small }

\maketitle

\begin{abstract}
Three coupled Ginzburg-Landau equations for 
hexagonal patterns with broken chiral symmetry are investigated.  
They are relevant for the dynamics close to onset of rotating 
non-Boussinesq or surface-tension-driven convection. Steady and 
oscillatory, long- and short-wave instabilities of the hexagons are 
found. For the long-wave behavior coupled phase equations are derived. 
Numerical simulations of the Ginzburg-Landau equations indicate bistability between spatio-temporally chaotic patterns and stable steady 
hexagons. The chaotic state can, however, not be described properly 
with the Ginzburg-Landau equations.\\

\\


\end{abstract}

\section{Introduction}

Convection has played a key role in the elucidation of the 
spatio-temporal dynamics arising in nonequilibrium pattern forming 
systems.  The interplay of well-controlled experiments with analytical 
and numerical theoretical work has contributed to a better 
understanding of various mechanisms that can lead to complex behavior.  
From a theoretical point of view the effect of rotation on roll 
convection has been particularly interesting because it can lead to 
spatio-temporal chaos immediately above threshold where the small 
amplitude of the pattern allows a simplified treatment.  Early work of 
K\"{u}ppers and Lortz \cite{KuLo69,Ku70} showed that for sufficiently 
large rotation rate the roll pattern becomes unstable to another set 
of rolls rotated with respect to the initial one.  Due to isotropy the 
new set of rolls is also unstable and persistent dynamics are 
expected.  Later Busse and Heikes \cite{BuHe80} confirmed experimentally the 
existence of this instability and the persistent dynamics arising from 
it.  They proposed an idealized model of three coupled 
amplitude equations 
%(nowadays known as the Busse-Heikes model)
in which the instability leads to a heteroclinic cycle connecting 
three sets of rolls rotated by 120$^o$ with respect to each other.  
Recently the K\"uppers-Lortz instability and the ensuing dynamics have 
been subject to intensive research, both experimentally 
\cite{ZhEc91,ZhEc92,NiEc93,HuEc95,HuPe98} and theoretically 
\cite{XiGu94,TuCr92,FaFr92,NeFr93,ClKn93,CrMe94,PoPa97}.  It is 
found that in sufficiently large systems the switching between rolls 
of different orientation looses coherence and the pattern breaks up 
into patches in which the rolls change orientation at different times.  The shape and 
size of the patches changes persistently due to the motion of the 
fronts separating them.  Other interesting aspects induced by rotation 
are the modification of the dynamics of defects \cite{MiBe95} and an 
unexpected transition to square patterns \cite{BaLi98}.

In this paper we are interested in the effect of rotation on hexagonal 
rather than roll (stripe) patterns as they arise in systems with 
broken up-down symmetry (e.g.  non-Boussines convection or 
surface-tension driven convection).  Complex dynamics, if they are 
indeed induced by the rotation, are likely to differ qualitatively 
from those in roll patterns due to the difference in the symmetry of 
the pattern.  Considering the small-amplitude regime close to onset, 
we use coupled Ginzburg-Landau equations.  On this level the Coriolis 
force arising from rotation manifests itself as a breaking of the 
chiral symmetry.  We therefore consider quite generally the effect of 
chiral symmetry breaking on weakly nonlinear hexagonal patterns.  
Since the equations are derived from the symmetries of the system we expect 
them to capture the generic behavior close to onset. 

 The 
dynamics of strictly periodic hexagon patterns with broken chiral 
symmetry have been investigated 
in detail by Swift \cite{Sw84} and Soward \cite{So85}.  They found 
that the heteroclinic orbit of the Busse-Heikes model is replaced by a 
periodic orbit arising from a secondary Hopf bifurcation off the 
hexagons.  Their results have been confirmed in numerical simulations 
of a Swift-Hohenberg-type model \cite{MiPe92}.  The competition 
between hexagons, rolls, and squares in rotating B\'enard-Marangoni 
convection has been considered in \cite{Ri94}.  In the present paper 
we focus on the impact of rotation on the side-band instabilities of 
steady hexagon patterns, i.e.  on instabilities that introduce modes 
with wavelengths or orientation different than those of the hexagons 
themselves. Thus, we extend  the work of Sushchik and Tsimring 
\cite{SuTs94} to 
the case of broken chiral symmetry. We find that rotation can {\it increase} the 
wavenumber range over which the hexagons are stable with respect to 
long-wave perturbations.  For larger values of the control parameter, 
however, additional short-wave instabilities arise.  The long- and the 
short-wave modes can be steady or oscillatory.  While in most cases they eventually lead 
to stable hexagon or roll patterns with different wavevectors, they 
can also induce persistent dynamics that can apparently not be 
described with Ginzburg-Landau equations.
 
The paper is organized as follows.  In the following section we use 
symmetry arguments to introduce the appropriate Ginzburg-Landau equations.  The stability with respect to long-wave 
perturbations is addressed in section III in which the coupled phase 
equations for the system are derived.  General perturbations (within 
the Ginzburg-Landau framework) are considered in section IV.  In section V we 
investigate numerically the nonlinear behavior resulting from the 
side-band instabilities.  Conclusions are given in section VI.

\section{Amplitude equations}

We consider small-amplitude hexagon patterns in systems with broken 
chiral symmetry.  For strictly periodic patterns the amplitudes ${\cal 
A}_i$ of the three sets of rolls (stripes) that make up the hexagon 
satisfy then the equations \cite{GoSw84}
\begin{equation}
\label{eq.amp1}
%\tau_0
\partial _{t}{\cal A}_{1}=\epsilon {\cal A}_{1}+\alpha_0\overline{\cal 
A}_{2}\overline{\cal A}_{3}-g_1 {\cal A}_{1}|{\cal 
A}_{1}|^{2}-g_{2}{\cal A}_{1}|{\cal A}_{2}|^{2}-g_{3}{\cal 
A}_{1}|{\cal A}_{3}|^{2},
\end{equation}
where the equations for the other two amplitudes are obtained by 
cyclic permutation of the indices and $\epsilon$ is a small parameter 
related to the distance from threshold. The overbar represents complex conjugation. These equations can be 
obtained from the corresponding physical equations (e.g. 
Navier-Stokes) using a 
perturbative technique with the usual scalings (\( {\cal A}\sim 
{\mathcal{O}}(\epsilon^{1/2} ) \) and \( \partial _{t}\sim 
{\mathcal{O}}(\epsilon) \)).  In order for all terms in 
(\ref{eq.amp1}) to be of the same order the coefficient of the 
quadratic term must be small, \( \alpha_0 \sim 
{\mathcal{O}}(\epsilon^{1/2} ) \).  This term arises from a resonance 
of the wavevectors of the three modes in the plane.  The broken chiral 
symmetry  manifests itself by the cross-coupling coefficients not being
equal, \( g_{2}\neq g_{3} \).

For completeness it should be noted that rotation leads in 
convection not only to a chiral symmetry breaking but 
also to a (weak) breaking of the translation symmetry due to the 
centrifugal force. In the following we will 
consider it to be negligible.  In addition, for 
sufficiently small Prandtl number rotation can render the primary 
instability oscillatory \cite{Ch61,ClKn93}.

\begin{figure}
\centerline{\epsfxsize=12cm \epsfbox{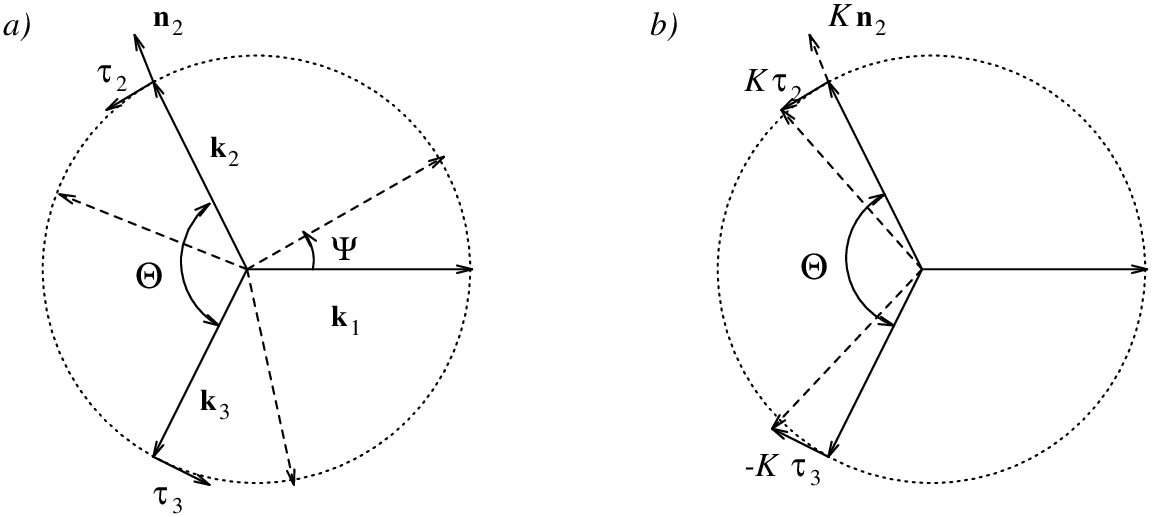}}
\caption{a) Critical wavenumbers \( {\bf k}_{1},{\bf k}_{2},{\bf k}_{3} \).
They satisfy the relation \( {\bf k}_{1}+{\bf k}_{2}+{\bf k}_{3}=0 \).
Up to a global rotation \protect\( \Psi \protect \), the position of the
wavevectors is given by the modulus of two of them and the angle between them.
The symbols \( {\bf n}_{i} \) and \( \mbox{\boldmath $\tau$}_{i} \)
denote the unit vectors in the parallel and perpendicular directions to the
wavevectors, respectively. b) The replacement
 \( A_{i}\rightarrow A_{i}e^{iK{\bf n}_{i}} \)
represents a change in the magnitude of the wavevector
 while \( A_{i}\rightarrow A_{i}e^{iK\mbox{\boldmath $\tau$}_{i}} \)
represents a rotation of it. \label{fig.ks} }
\end{figure}

In order to analyze the possibility of modulational instabilities 
spatial derivatives must be included in Eq.  (\ref{eq.amp1}).  We take 
the gradients in both directions to be of the same order, \( \nabla 
\sim {\mathcal{O}}(\epsilon^{1/2} ) \) \cite{Po86}, and retain both 
linear and quadratic gradient terms.  After rescaling the amplitude, 
time, and space we arrive at the equations,
\begin{eqnarray}
\partial _{t}A_{1} & = & \mu A_{1}+({\bf {n}}_{1}\cdot \nabla 
)^{2}A_{1}+\overline{A}_{2}\overline{A}_{3} - A_{1}|A_{1}|^{2}-(\nu 
+\tilde{\nu} )A_{1}|A_{2}|^{2}-(\nu -\tilde{\nu} 
)A_{1}|A_{3}|^{2}\nonumber \\
& & + i(\alpha _{1}+\tilde{\alpha} )\overline{A}_{2}({\bf 
{n}}_{3}\cdot \nabla )\overline{A}_{3}+i(\alpha _{1}-\tilde{\alpha} 
)\overline{A}_{3}({\bf {n}}_{2}\cdot \nabla )\overline{A}_{2}\nonumber 
\\
& & + i\alpha _{2}\left(\overline{A}_{2}(\mbox{\boldmath $\tau $}_{3}\cdot 
\nabla )\overline{A}_{3}-\overline{A}_{3}(\mbox{\boldmath $\tau 
$}_{2}\cdot \nabla )\overline{A}_{2}\right) \label{eq.ampgen}
\end{eqnarray}
where now all the coefficients are ${\cal O}(1)$, and ${\bf n}_i$ and 
$\mbox{\boldmath $\tau$}_i$ represent the unit vectors parallel and 
perpendicular to the wavenumber ${\bf k}_i$ (Fig.  \ref{fig.ks}).  The 
cross-coupling coefficients have been rewritten in terms of $\nu$ and 
$\tilde{\nu}$, with $\tilde{\nu}$ being proportional to rotation and 
therefore giving a measure of the chiral symmetry breaking. In the 
gradient terms the chiral symmetry breaking manifests itself in the 
terms proportional to \( \tilde{\alpha} \).

The influence of the nonlinear gradient terms in (\ref{eq.ampgen}) 
involving $\alpha_{1}$ and $\alpha_{2}$ has been studied by several 
authors \cite{Br89,GuOu94,EcPe98}.  The origin of the new term involving 
$\tilde{\alpha}$ is best understood by considering the coefficient 
$\alpha$ of the quadratic term.  The gradient terms arise from its 
dependence on the wavenumber of the modes involved, \( \alpha 
=\alpha ({\bf {k}}_{1},{\bf {k}}_{2},{\bf {k}}_{3}) \) i.e.  when the 
equation is considered in Fourier space.  Due to the 
resonance condition (\( {\bf {k}}_{1}+{\bf {k}}_{2}+{\bf {k}}_{3}=0 \)) 
we can drop the dependence on one of the wavevectors. Then, up to an 
arbitrary global rotation \( \Psi \), the system can be specified 
by the angle 
\( \Theta \) between the wavevectors ${\bf k}_2$ and ${\bf k}_3$ and 
their moduli \( k_{2} \) and \( k_{3} \) (see Fig.  \ref{fig.ks}). 
 Due to isotropy 
the coefficient \( \alpha \) cannot depend on the global rotation and 
can therefore be expressed as \( \alpha =\alpha (k_{2},k_{3},\Theta ) 
\).  When evaluated at the critical values of the wavenumbers $\alpha_0$ is given by 
$\alpha_0=\alpha(k^c_{2},k^c_{3},\Theta^c=2\pi/3)$ (in Eq.  
(\ref{eq.ampgen}) we take the normalization $\alpha_0=1$).  Since
a change in the modulus of \( {\bf k}_{i} \)  
can be effected by the replacement  \( A_{i}\rightarrow A_{i}e^{iK{\bf n}_{i}} \)
the dependence of $\alpha$ on the moduli $k_{2}$ and $k_{3}$ can be 
represented  in real space  by
\begin{equation}
\frac{\partial \alpha }{\partial k_{2}}\overline{A}_{3}({\bf {n}}_{2}\cdot \nabla )\overline{A}_{2} + \frac{\partial \alpha }{\partial k_{3}}\overline{A}_{2}({\bf {n}}_{3}\cdot \nabla )\overline{A}_{3}.
\end{equation}
When the chiral symmetry is broken $ \partial \alpha /\partial k_{2}$ 
and $\partial \alpha /\partial k_{3} $ need not be equal and it is 
convenient to introduce the coefficients
\begin{equation}
\alpha_1=\frac{1}{2}\left ( \frac{\partial \alpha}{\partial k_3} + \frac{\partial \alpha}{\partial k_2} \right ) \makebox[1cm]{and} 
\tilde{\alpha}=\frac{1}{2}\left ( \frac{\partial \alpha}{\partial k_3} - 
\frac{\partial \alpha}{\partial k_2} \right )
\end{equation}
with $\alpha_{1}$ even and $\tilde{\alpha}$ odd in the amplitude of 
the symmetry breaking.

On the other hand, a variation in the angle between \( {\bf {k}}_{2} \)
and \( {\bf {k}}_{3} \) is represented by 
\begin{equation}
\frac{i}{k_c}\frac{\partial \alpha }{\partial \Theta }(\overline{A}_{2}(\mbox{\boldmath $\tau$}_{3}\cdot \nabla )\overline{A}_{3}-\overline{A}_{3}(\mbox{\boldmath $\tau$}_{2}\cdot \nabla )\overline{A}_{2})
\end{equation}
with only one coefficient \( \alpha _{2}=(\partial \alpha /\partial 
\Theta)/k_c \).  This term is invariant under reflections 
interchanging modes $A_2$ and $A_3$, since in contrast to the normal 
vector ${\bf n}_i$ the tangential vector $\mbox{\boldmath $\tau$}_i$ 
changes sign in this reflection. Therefore the coefficient $\alpha_2$ 
is even in the amplitude of the symmetry breaking.
%\leftrightarrow -{\bf tau}_3$ \( k_{2}\leftrightarrow k_{3} \).

\begin{figure}
\centerline{\epsfysize=5cm\epsfbox{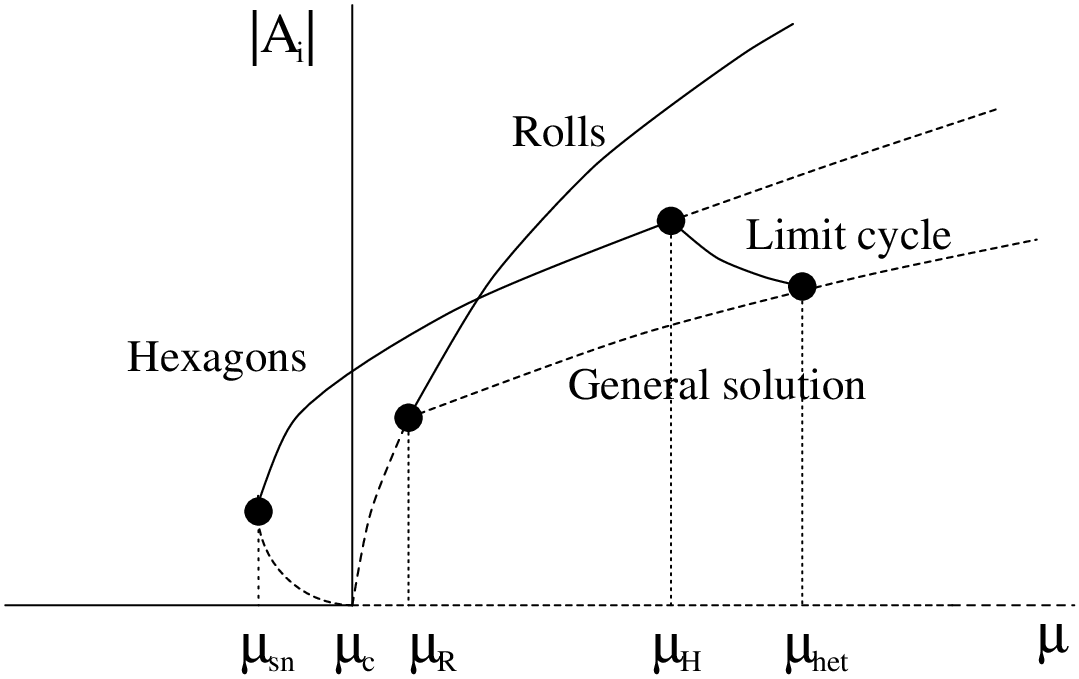}}

\caption{Sketch of a bifurcation diagram for a fixed value of the wavenumber $K$. \label{fig.bifos}}
\end{figure}

Equation (\ref{eq.ampgen}) admits hexagon solutions 
 \( A_{i}=He^{iK\hat{\bf n}_{i}\cdot {\bf {x}}+i\phi_i} \)  with a sligthly offcritical 
wavenumber (${\bf k}_i={\bf k}^c_i + {\bf K}_i$, $K \ll k_c$), with
\begin{equation}
H=\frac{(1+2K\alpha _{1})\pm \sqrt{(1+2K\alpha _{1})^{2}+4(\mu -K^{2})(1+2\nu )}}{2(1+2\nu )}, \qquad \Phi\equiv \phi_1+\phi_2+\phi_3=0.
\end{equation}
The stability of this solution to   perturbations with the same 
wavevectors has been studied by 
several authors \cite{Sw84,So85} and can be summarized in the 
bifurcation diagram shown in Fig.  \ref{fig.bifos}.  The hexagons appear 
through a saddle-node bifurcation at $\mu=\mu_{sn}$,
\begin{equation}
\mu_{sn} = -\frac{(1+2K\alpha_1)^2}{4(1+2\nu)} + K^2,
\end{equation}
 and become unstable $via$ a Hopf bifurcation at \( \mu =\mu _{H} \),
\begin{equation}
\mu_H = \frac{(1+2K\alpha_1)^2 (2+\nu)}{(\nu-1)^2} + K^2,
\end{equation}
with a critical frequency \( \omega _{c}=2\sqrt{3}\tilde{\nu}(1+2K\alpha_1)^2 /(\nu -1)^{2} \).  Note that the Hopf frequency does not depend on 
$\tilde{\alpha}$. The Hopf bifurcation is supercritical and for $\mu > 
\mu_H$  stable oscillations in the three amplitudes of the 
hexagonal pattern arise with a phase shift of $2\pi/3$ between them 
\cite{Sw84,So85,MiPe92}, resulting in what we are going to call 
oscillating hexagons.  As  \( \mu \) is increased further, eventually a 
point $\mu=\mu_{het}$ is reached at which the branch of oscillating hexagons ends on the branch corresponding to the mixed-mode solution in a global bifurcation involving a heteroclinic connection.  Above this point the only stable 
solution is the roll solution whose stability region is bounded below by
\begin{equation}
\mu _{R}=\frac{1}{(\nu +\tilde{\nu} -1)(\nu -\tilde{\nu} -1)}+K^{2}.
\end{equation}
When $|\tilde{\nu}| > \nu-1$ the rolls are never stable and the 
limit cycle persists for arbitrary large values of $\mu$. In the absence 
of the quadratic terms in Eq. (\ref{eq.amp1}) this condition corresponds to the 
K\"uppers-Lortz instability of rolls.

\section{Long-Wave approximation: Phase equation}

Already below the Hopf bifurcation to oscillating hexagons the hexagonal 
pattern can be unstable to side-band perturbations.  
The behavior of long-wavelength modulations is described by the 
dynamics of the phase of the periodic structure. The phases \( 
\phi _{i} \) of the three modes of the hexagonal pattern can be 
combined to define 
a phase vector \( \mbox{\boldmath $\phi$}\equiv (\phi_x,\phi_y)\), the 
components \(\phi _{x}\equiv -(\phi _{2}+\phi _{3})\) and \(\phi 
_{y}\equiv (\phi _{2}-\phi _{3})/\sqrt{3} \) of which are related to 
the translation modes in the \( x \) - and the \( y \) -direction.  In 
the chirally symmetric case the phase vector satisfies the coupled 
diffusion equations \cite{LaMe93,Ho95}
\begin{equation}
\tau _{0}\partial _{t}\mbox{\boldmath $\phi$ }=D_{\bot }\nabla 
^{2}\mbox{\boldmath $\phi$ }+(D_{\Vert }-D_{\bot })\nabla (\nabla 
\cdot \mbox{\boldmath $\phi$}),
\end{equation}
and can be decomposed into a longitudinal (irrotational) and a 
transversal (divergence free) part \( \mbox{\boldmath $\phi$} \equiv 
\nabla \psi _{l}+\nabla \times \hat{\bf e}_{z}\psi _{t} \).  The fields 
$\psi_{l,t}$ each satisfy a diffusion equation with diffusion constants \( 
D_{\Vert } \) and \( D_{\bot } \), respectively.

We can use symmetry arguments to derive the form of the phase equation 
when the chiral symmetry is broken.  We consider a general diffusion 
coefficient which is a tensor of rank four,
\begin{equation}
\partial _{t}\phi _{i}=D_{i}^{jkl}\partial _{j}\partial _{k}\phi _{l}
\end{equation}
with summation over repeated indices implied.  Invariance under 
reflection and rotations of \( 60^{o} \) restrict the number of 
possible independent coefficients $D_{i}^{jkl}$.  In the case of broken chiral 
symmetry we split the diffusion tensor into two parts: one even under 
reflections, the other odd,
\begin{equation}
D_{i}^{jkl}=\overline{D}_{i}^{jkl}+\Omega _{i}^{m}\tilde{D}_{m}^{jkl},
\end{equation}
 where \( \Omega _{i}^{m} \) is the antisymmetric tensor of rank two given
by 
\begin{equation}
\Omega =\left( \begin{array}{cc}
0 & -\omega \\
\omega  & 0
\end{array}\right),
\end{equation}
 with \( \omega  \) giving the strength of the chiral symmetry breaking (e.g.
the rotation frequency). \( \bar{D} \) and \( \tilde{D} \) are even functions
of \( \omega  \). As generators of the symmetry group  we can take rotations
of \( 60^{o} \) and reflections in \( x \),
\begin{equation}
R_{60}=\left( \begin{array}{cc}
\frac{1}{2} & \frac{\sqrt{3}}{2}\\
-\frac{\sqrt{3}}{2} & \frac{1}{2}
\end{array}\right) ,\; \; \kappa_{x}=\left( \begin{array}{cc}
-1 & 0\\
0 & 1
\end{array}\right).\label{e:sym}
\end{equation}
Requiring that $\bar{D}$ and $\tilde{D}$ are invariant under the operations (\ref{e:sym}) 
one can show that the most general form of the phase equation
with broken quiral symmetry is given by
\begin{equation}
\label{eq.phaserot}
\partial _{t}\mbox{\boldmath $\phi$}=D_{\bot }\nabla ^{2}\mbox{\boldmath $\phi$}+
(D_{\Vert }-D_{\bot })\nabla (\nabla \cdot \mbox{\boldmath $\phi$})-
D_{\times _{1}}(\hat{\bf e}_{z}\times \nabla ^{2} \mbox{\boldmath $\phi$})+
D_{\times _{2}}(\hat{\bf e}_{z}\times \nabla )(\nabla \cdot \mbox{\boldmath 
$\phi$}),
\end{equation}
 where \( \hat{\bf e}_{z} \) is a unit vector in the direction perpendicular to
the plane. 

It is worth  emphasizing that, although the coefficients
of this equation can be derived from the amplitude equations, its form is given
by symmetry arguments and is, therefore, generic and valid even far from threshold.
To  derive the phase equation from the amplitude
equations (\ref{eq.ampgen}) we consider a perfect hexagonal pattern
with a wavenumber slightly different from critical (\( k=k_{c}+K \)) and perturb
it, both in amplitude and phase, $A_{i}=(H+r_{i})e^{iK\hat{\bf n}_{i}\cdot {\bf {x}}+i\phi _{i}}$. Away from threshold, from
the saddle-node and the Hopf bifurcation,  the amplitude modes \( r_{1},r_{2},r_{3} \),
and the global phase \( \Phi =\phi _{1}+\phi _{2}+\phi _{3} \) are strongly 
damped and can be eliminated adiabatically. Following the usual procedures 
(e.g. \cite{Ma90}) we 
arrive at Eq. (\ref{eq.phaserot}) with
\begin{eqnarray}
D_{\bot } & = & \frac{1}{4}+\frac{1}{u^{2}+\omega^{2}}
\left\{\frac{1}{4}H^{2}u[(\alpha_{1}+\sqrt{3}\alpha_{2})^{2}+3\tilde{\alpha}^2]
- \sqrt{3}H\omega \tilde{\alpha} K - u K^2 \right\},\\
D_{\Vert } & = & D_{\bot }+\frac{1}{2}-
\frac{1}{v}\left\{H^{2}\alpha_{1}(\alpha _{1}-\sqrt{3}\alpha _{2})-
H (3\alpha_{1}-\sqrt{3}\alpha _{2})K+2K^{2}\right\},\\
D_{\times _{1}} & = & \frac{1}{u^2+\omega ^{2}}
\left\{\frac{1}{4} \omega H^2[(\alpha_{1}+\sqrt{3}\alpha _{2})^{2}+3\tilde{\alpha} 
^{2}]+\sqrt{3}Hu\tilde{\alpha}K- \omega K^2\right\} ,\\
D_{\times _{2}} & = & \frac{\tilde{\alpha}}{v}
\left\{\sqrt{3}H^{2}\alpha_{1} - \sqrt{3}HK \right\}.
\end{eqnarray}
where
\bea 
&&\omega =2\sqrt{3}H^{2}\tilde{\nu},\\
 &&u=2H^{2}(1-\nu 
)+2(1+2K\alpha _{1})H,\\
 &&v=2H^{2}(1+2\nu )-(1+2K\alpha _{1})H. 
\eea
The coefficients \( D_{\times _{1}} \) and \( D_{\times _{2}} \) are 
odd in the symmetry-breaking terms $\tilde{\nu}$ and $\tilde{\alpha}$.
At the Hopf bifurcation curve \( u=0 \) implying 
$H=(1+2K\alpha_1)/(\nu -1)$ and $\omega=\omega_c$, 
while \( v=0 \) represents the saddle-node instability.

\begin{figure}
\centerline{\epsfxsize=6cm\epsfbox{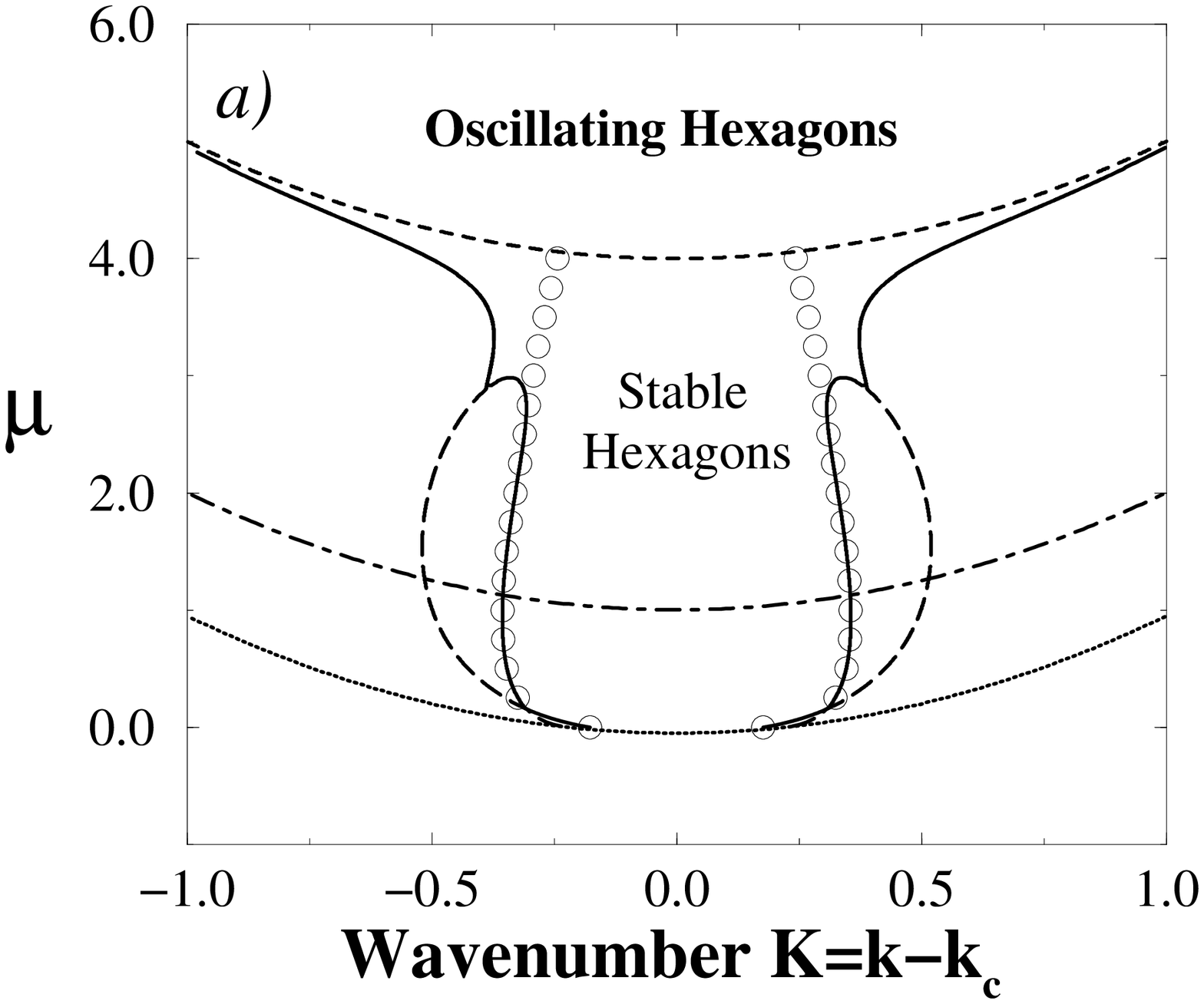}
\epsfxsize=6cm\epsfbox{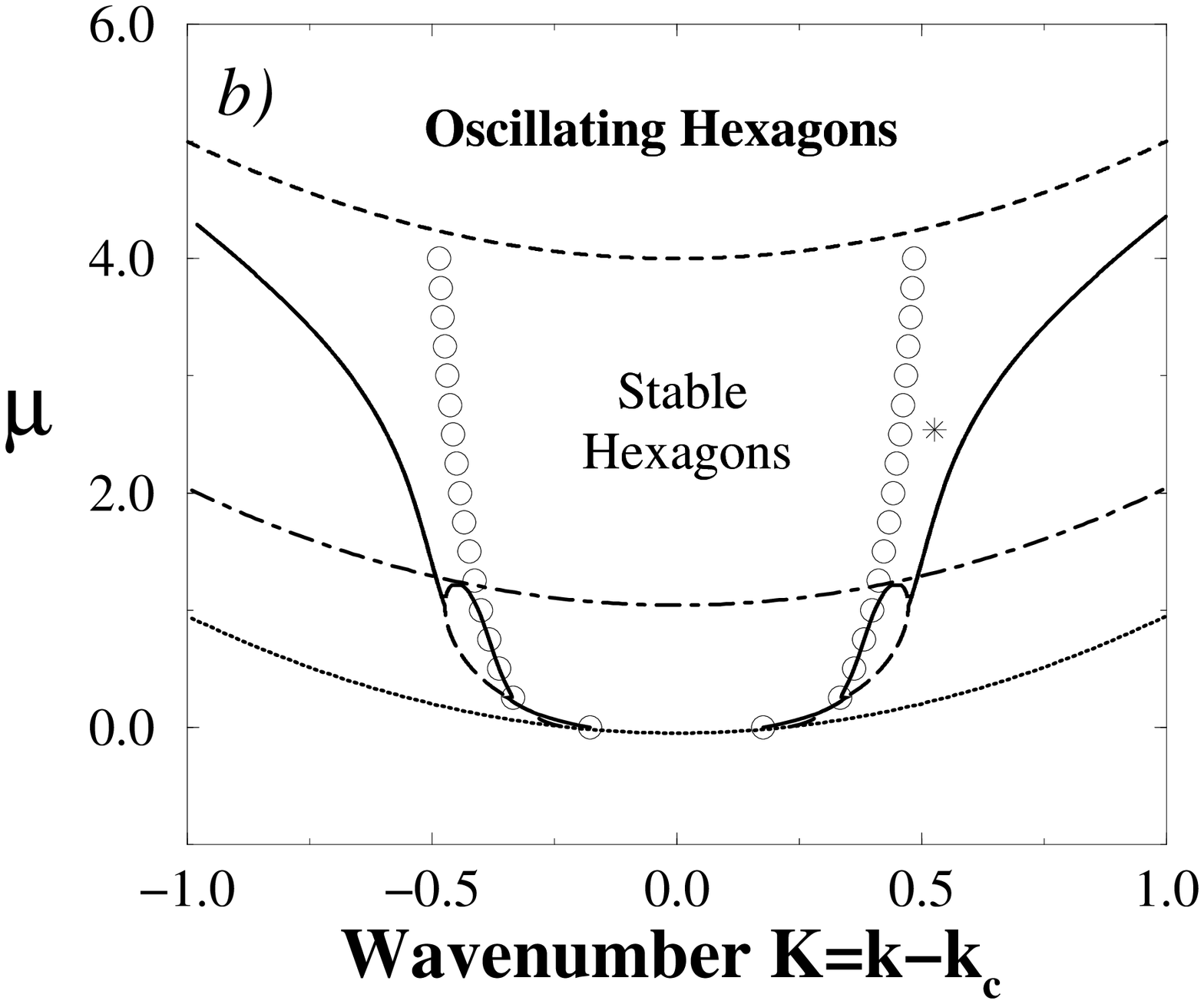}
\epsfxsize=6cm\epsfbox{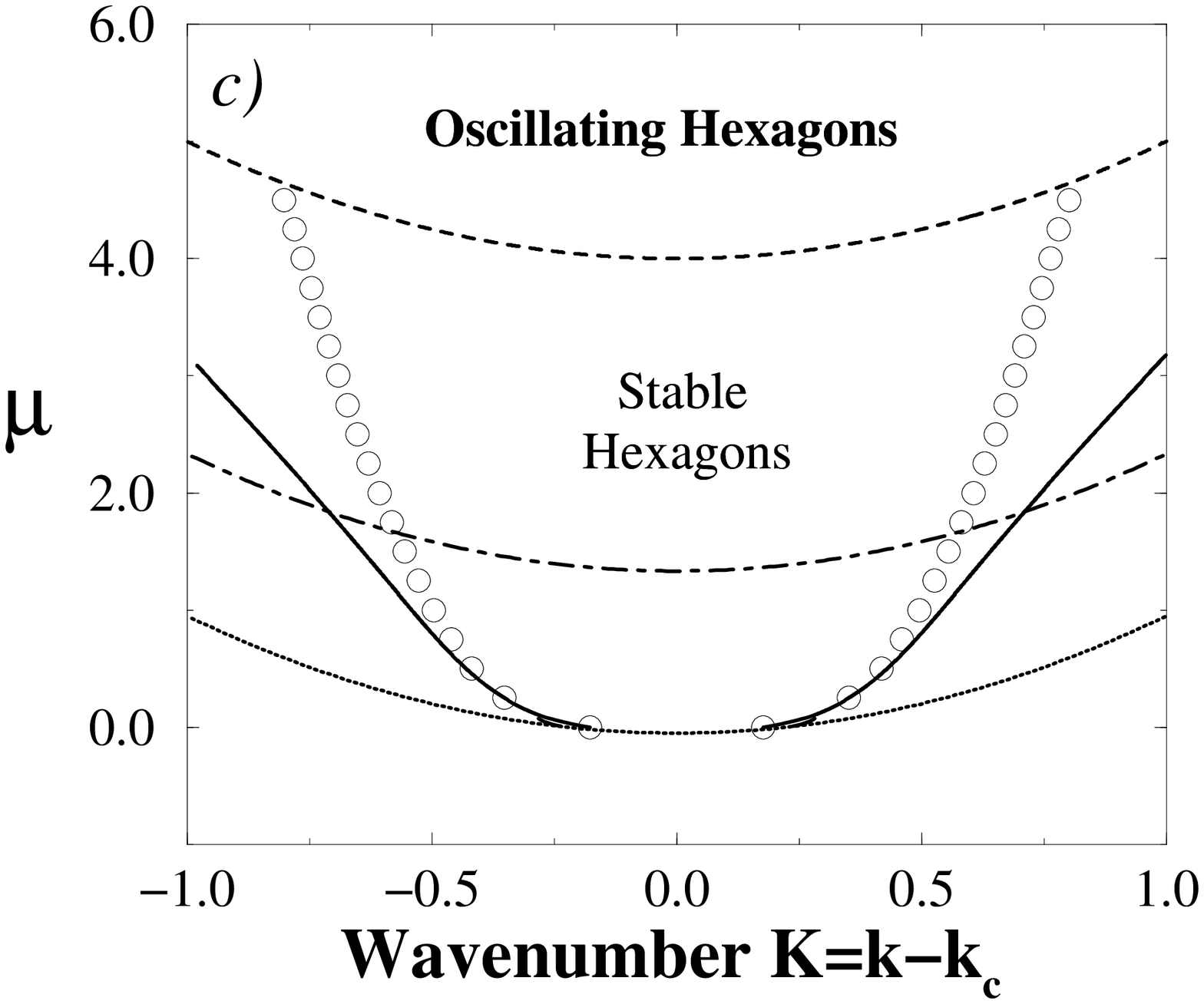}}

\caption{Stability diagrams for \protect\( \nu =2\protect \), 
\protect\( \tilde{\alpha} =\alpha _{1}=\alpha _{2}=0\protect \)
and a) \protect\( \tilde{\nu} =0.05\protect \), b) \protect\( \tilde{\nu} =0.2\protect \) 
and c) \protect\( \tilde{\nu} =0.5\protect \). 
The dotted and short-dashed lines indicate the saddle-node and the Hopf bifurcation to 
oscillating hexagons, respectively. 
The solid and dashed lines correspond to the long-wave 
instabilities, while the open circles are 
the results from the general stability analysis. Rolls are stable 
with respect to hexagons above the dot-dashed line.}

\label{fig.gam}
\end{figure}

Expanding the phase in normal modes \( \mbox{\boldmath $\phi$}=\mbox{\boldmath $\phi$}^{0}e^{i{\bf {Q}}\cdot {\bf {x}}+\sigma t} \)
we obtain the dispersion relation 
\begin{equation}
\sigma ^{2}+(D_{\Vert }+D_{\bot })Q^{2}\sigma +\left(D_{\Vert }D_{\bot 
}+D_{\times _{1}}(D_{\times _{1}}+D_{\times _{2}})\right)Q^{4}=0,
\end{equation}
 whose eigenvalues are 
\begin{equation}
\label{eq.disp}
\sigma _{1,2}=-\frac{1}{2}\left[ D_{\Vert }+
D_{\bot }\pm \sqrt{(D_{\Vert }-D_{\bot })^{2}-4D_{\times _{1}}(D_{\times
_{1}}+D_{\times _{2}})}\right] Q^{2}. \label{ec.sigma}\end{equation}
When \( D_{\times _{1}}=0 \) the eigenvalues become simply \( \sigma _{1}=-D_{\Vert }Q^{2} \)
and \( \sigma _{2}=-D_{\bot }Q^{2} \), corresponding to the eigenvalues of
the irrotational and the divergence-free phase modes, respectively. If the rotation rate is small (\( D_{\times _{1}},D_{\times _{2}}\ll D_{\bot },D_{\Vert } \))
we can expand (\ref{eq.disp}) and obtain,
\begin{equation}
\sigma _{1}=-\left (D_{\Vert }-\frac{D_{\times _{1}}(D_{\times 
_{1}}+D_{\times _{2}})}{(D_{\Vert }-D_{\bot })}\right )Q^{2},\qquad 
\sigma _{2}=-\left (D_{\bot }+\frac{D_{\times _{1}}(D_{\times 
_{1}}+D_{\times _{2}})}{(D_{\Vert }-D_{\bot })}\right )Q^{2}.
\end{equation}
For \( D_{\times _{1}},D_{\times _{2}}\sim D_{\Vert 
},D_{\bot } \) this approximation is not valid and the longitudinal 
and transverse perturbations become coupled.  An important novelty in 
this case is that the phase instability can become oscillatory. It 
occurs when the following conditions are satisfied,
\begin{eqnarray}
 &  & D_{\Vert }+D_{\bot }=0,\\
 &  & (D_{\Vert }-D_{\bot })^{2}-4D_{\times _{1}}(D_{\times _{1}}+D_{\times _{2}})=0.\label{ec.osci.fase} 
\end{eqnarray}
 In Fig. \ref{fig.gam} and Fig. \ref{fig.eta} (below) we represent the phase instability
curves for a number of cases. For small values of the rotation rate \( \tilde{\nu}  \),
the phase stability diagram is similar to that obtained in the absence of rotation,
especially for small \( \mu  \). As $|K|$ is increased, both real eigenvalues
in  (\ref{ec.sigma}) go through zero consecutively as indicated by the dashed
and solid lines in Fig.\ref{fig.gam}. As \( \mu \) is increased towards the
Hopf bifurcation (\( \mu \rightarrow \mu _{H} \)) the two lines merge and the
phase instability becomes oscillatory as indicated by the solid lines. Note
that, in contrast to the chirally symmetric case, the left and right stability
limits do not merge at \( K=0 \) as the transition to oscillating hexagons is
reached. Instead, they are open and over a range of wavenumbers the hexagons
remain stable with respect to long-wave perturbations all the way to the Hopf
bifurcation at \( \mu =\mu _{H} \).  Furthermore, while in the chirally
symmetric case the analog of the Hopf bifurcation is transcritical (and
steady) and leads discontinously to rolls, the Hopf bifurcation is
supercritical and leads to oscillating hexagons \cite{Sw84,So85,MiPe92}.

\section{General Stability Analysis}

\begin{figure}
{\par\centering \parbox{5.5cm}{\epsfxsize=5.5cm \centering\epsfbox{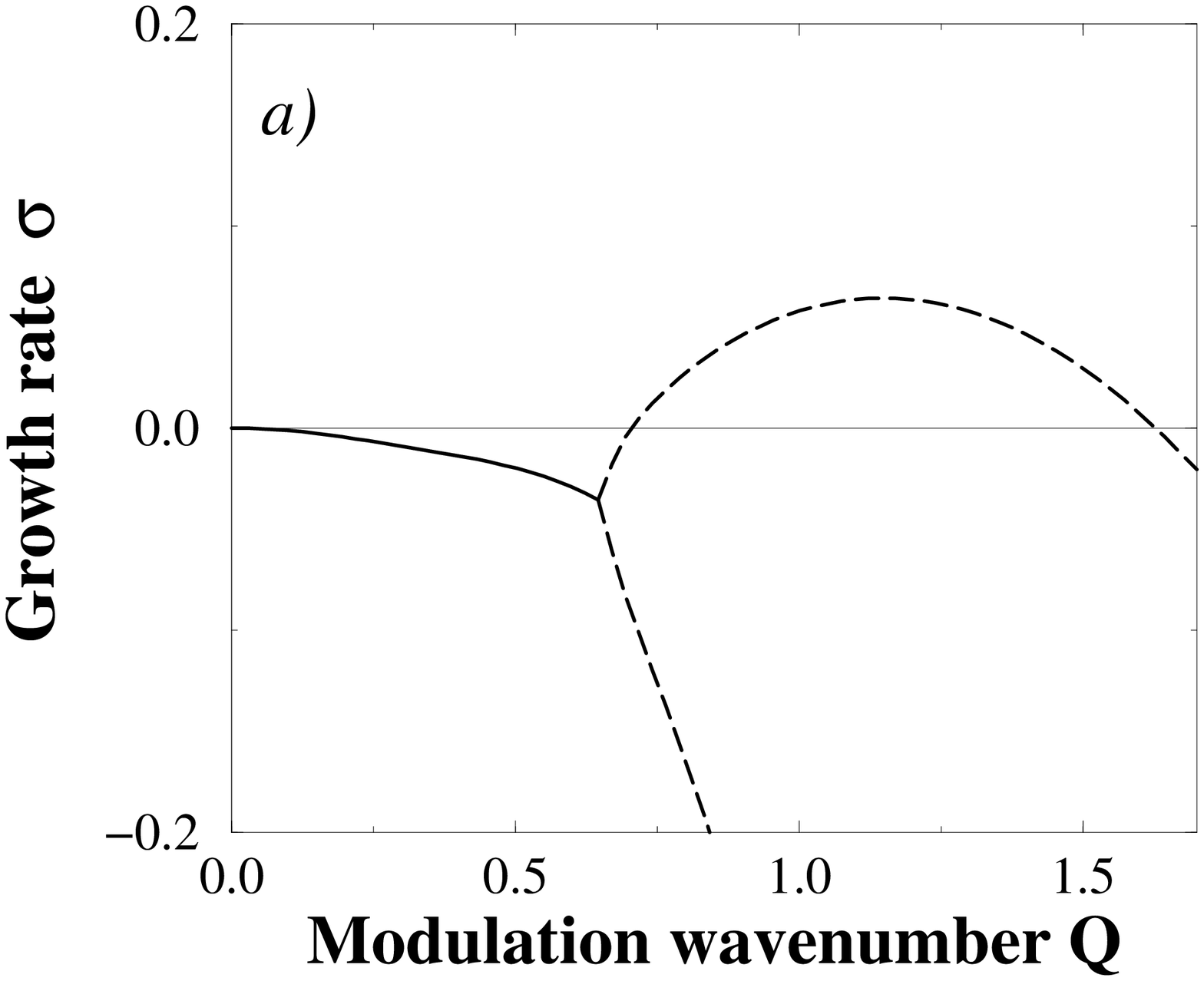}} \parbox{5.5cm}{\epsfxsize=5.5cm
\epsfbox{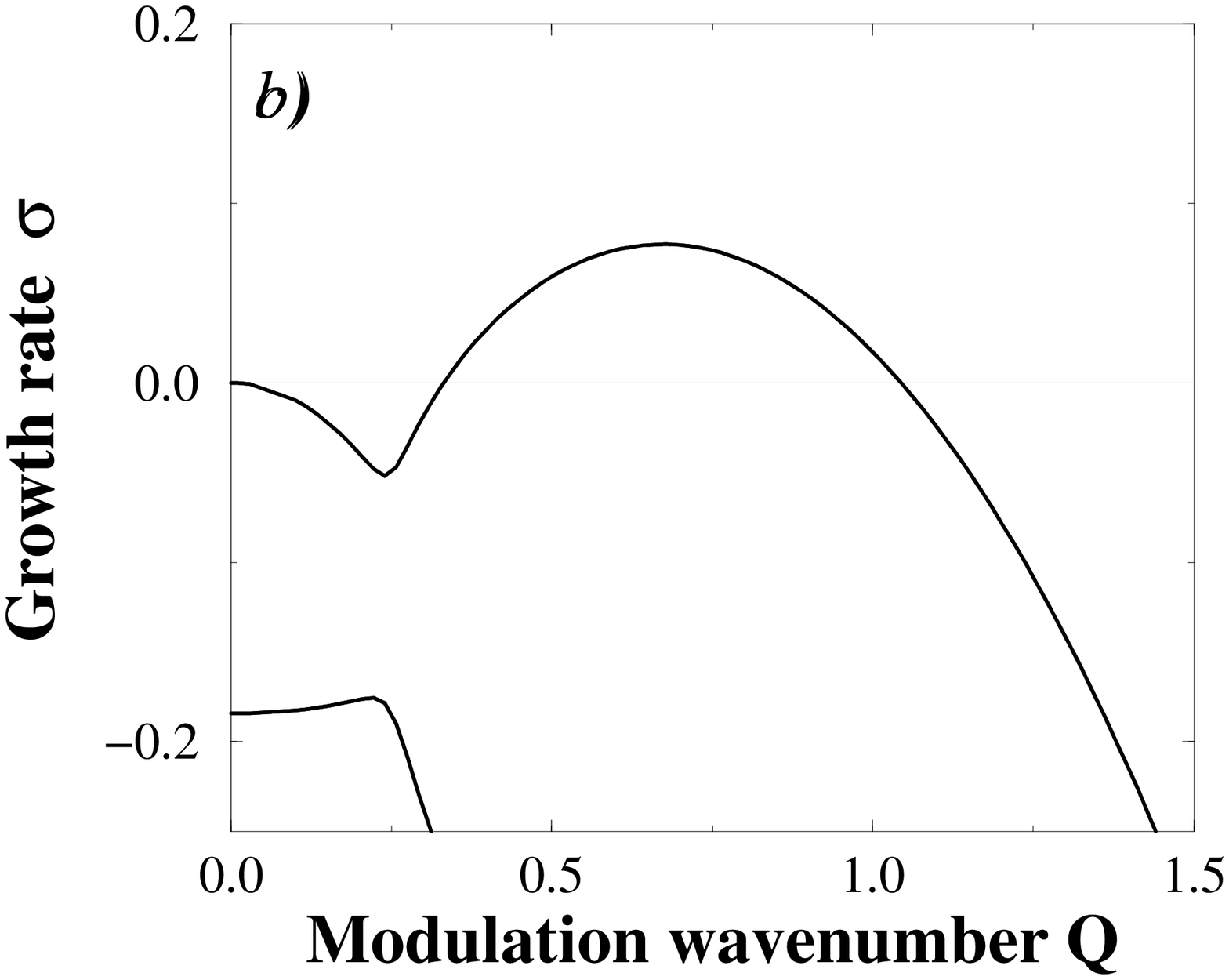}} \parbox{5.5cm}{\epsfxsize=5.5cm \epsfbox{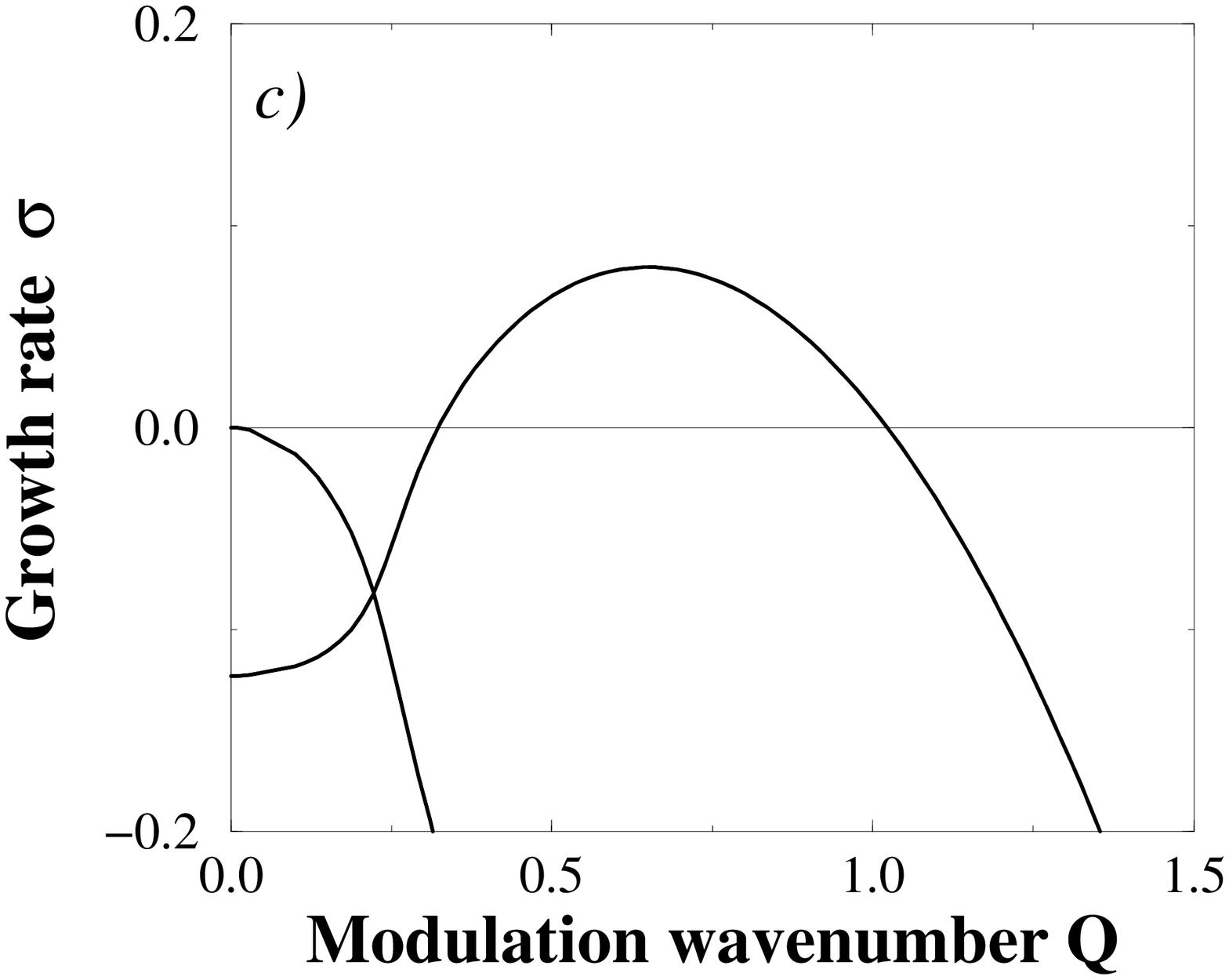}}\par}

\caption{Growth rates (with $\nu=2$, $\tilde{\nu}=0.1$, 
$\alpha_1=\alpha_2=0$) when a) the translation modes are short-wave 
unstable (\protect\( \tilde{\alpha} =0\protect \) and \protect\(\mu 
=2.54\protect \), $q=0.53$).  The Hopf modes interact with the translation 
modes and their eigenvalues become positive at finite wavenumber (\protect\( 
\tilde{\alpha} =0.4\protect \)): b) Just below the branch switching 
(\protect\( \mu =3.49\protect \), $q=-0.6$), and c) when the Hopf mode 
is unstable (\protect\( \mu =3.75\protect \), $q=-0.57$).  The solid 
and dashed lines correspond to complex and real eigenvalues.  The
points at which these dispersion relations are obtained are marked by an asterisk in 
Figs.  \ref{fig.gam} and \ref{fig.eta}.}

{\par\centering \label{fig.disp}\par}
\end{figure}

We now consider arbitrary perturbations of the hexagonal pattern 
$A_{i}=(H+a_{i}e^{i{\bf {Q}}\cdot {\bf {x}}+\sigma t})e^{iK\hat{\bf 
n}_{i}\cdot {\bf {x}}}$, with \( a_{1} \), \( a_{2} \), \( a_{3} \) 
complex and solve the resulting \( 6\times 6 \) linearized system.  Two of the six eigenvalues correspond to 
the global phase \( \Phi \) and the overall amplitude involved in the 
saddle-node bifurcation.  In the regime of interest both are strongly 
negative.  The next two correspond to the translation modes, and can 
be real or complex.  It turns out that these modes can destabilize the 
hexagons not only \emph{via} the longwave instabilities 
(\ref{eq.disp}) but also \emph{via} short-wave instabilities as 
illustrated in Fig.  \ref{fig.disp}a, where the solid and dashed lines 
correspond to the real parts of the complex and real eigenvalues, respectively.  Finally, there is a 
pair of complex conjugate eigenvalues corresponding to the Hopf 
bifurcation to oscillating hexagons.  For some parameter values these 
eigenvalues merge with the ones corresponding to the phase modes and 
their real parts become positive (Fig.  \ref{fig.disp}b and 
\ref{fig.disp}c).  Note that in the chirally symmetric case the 
eigenvalues are always real and the instabilities long-wave 
\cite{SuTs94}.

In what follows we will consider $\alpha_1=\alpha_2=0$ for simplicity.  
Although non-zero values for these coefficients change the stability boundaries 
quantitatively, they are not found to induce any qualitatively 
different instability.

\begin{figure}
\centerline{\epsfxsize=7cm\epsfbox{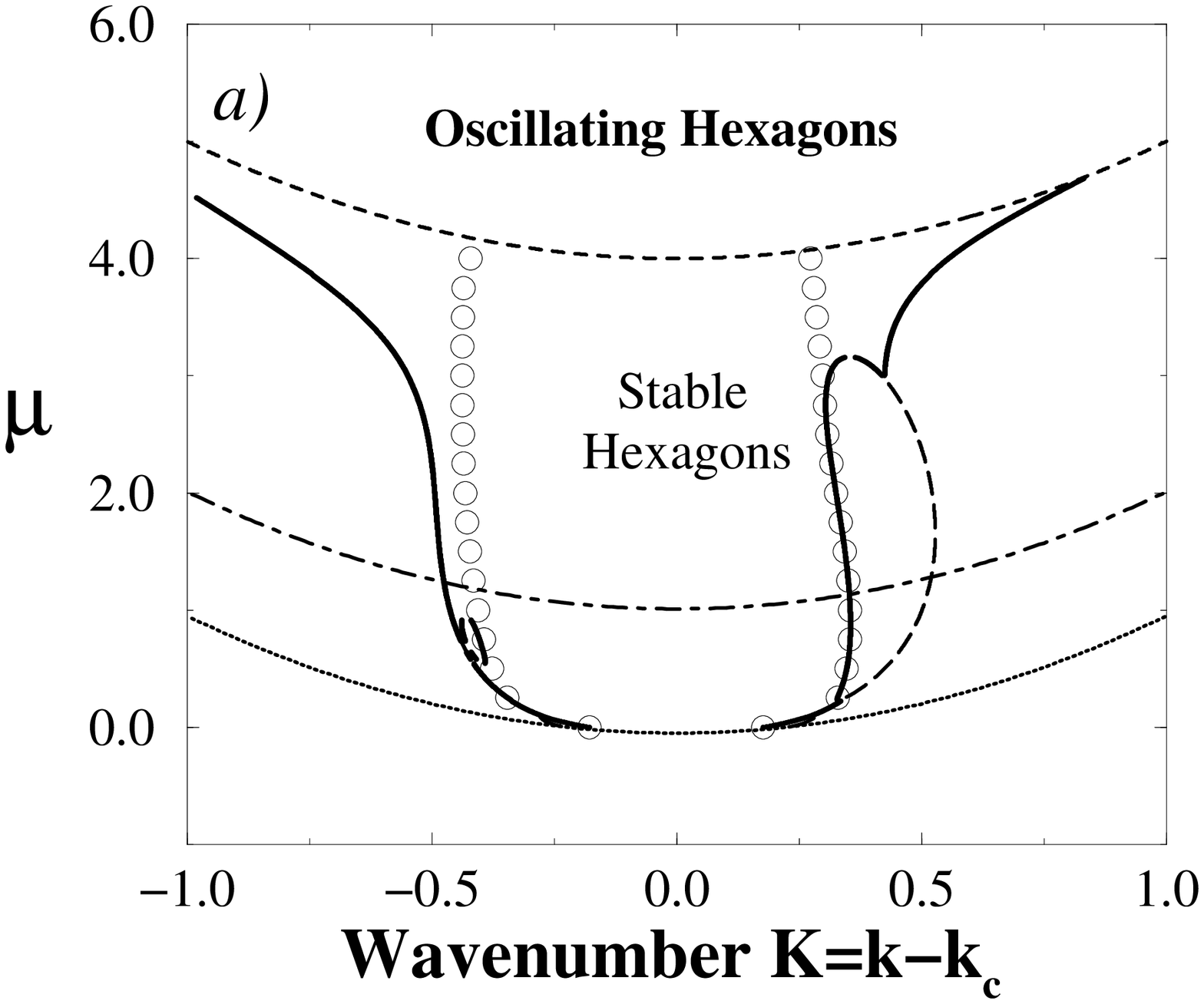}
\epsfxsize=7cm\epsfbox{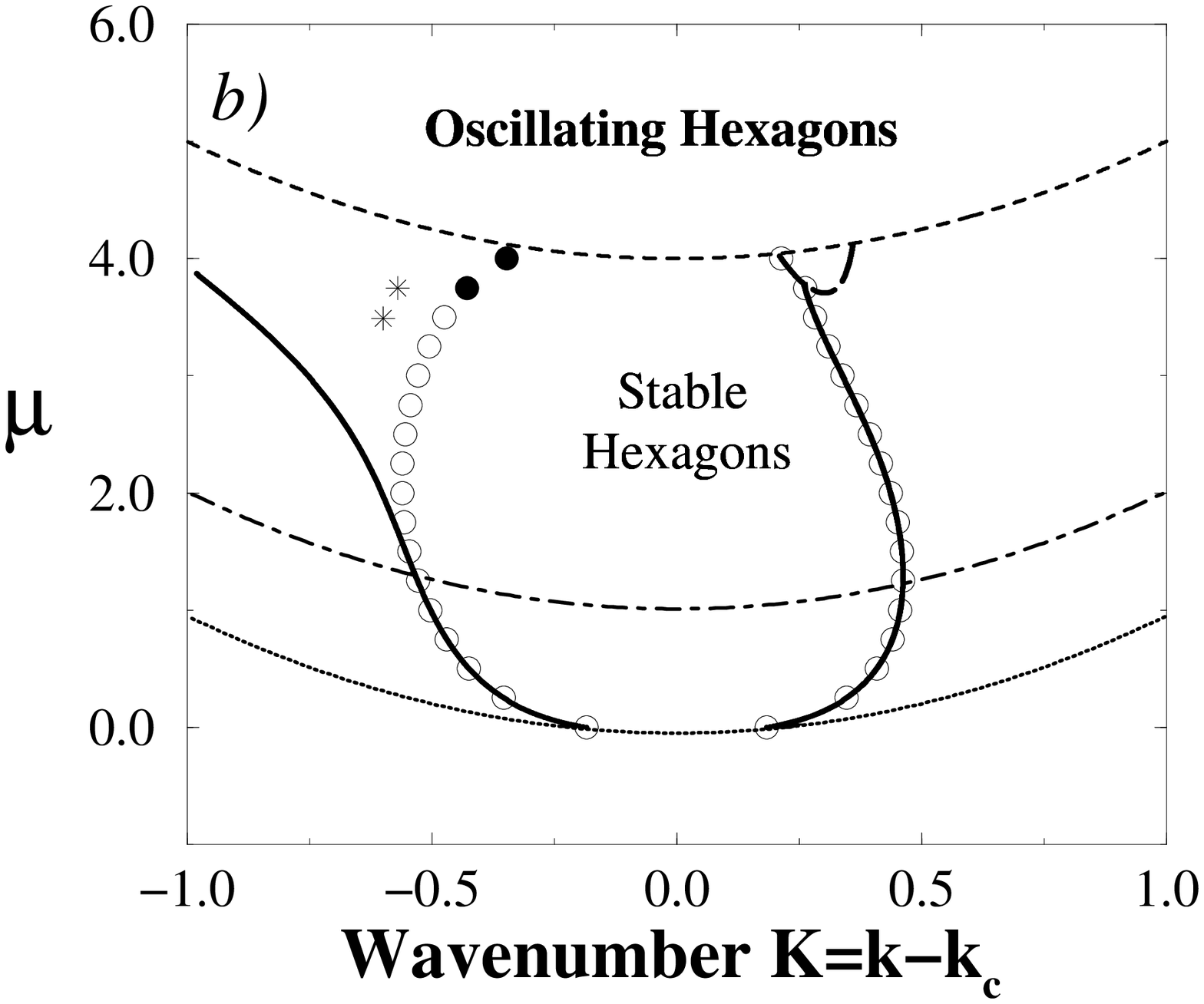}}
\centerline{\epsfxsize=7cm\epsfbox{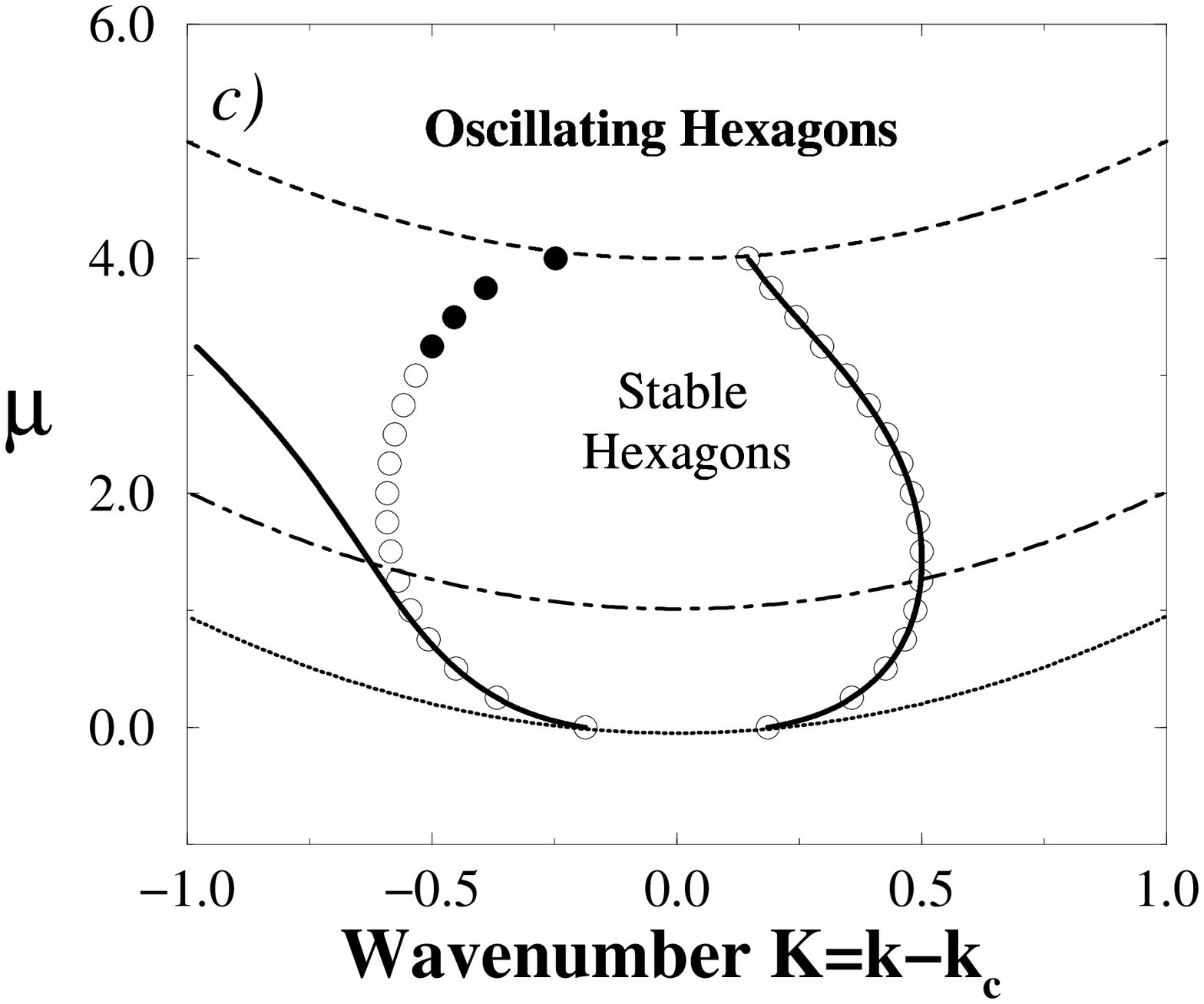}
\epsfxsize=7cm\epsfbox{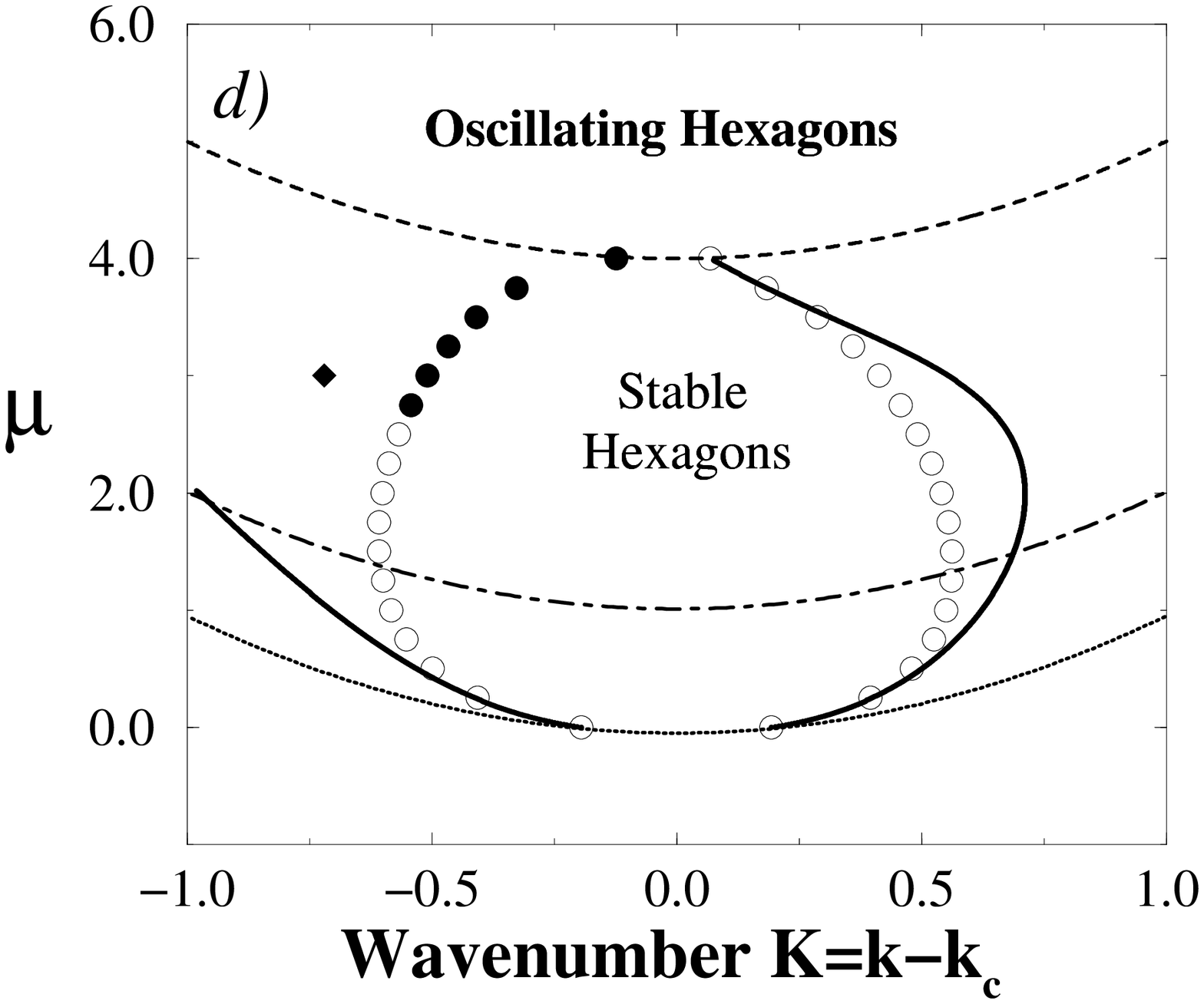}}

\caption{Stability diagrams for \protect\( \nu =2\protect \), \protect\( \tilde{\nu} =0.1\protect \),
\protect\( \alpha _{1}=\alpha _{2}=0\protect \) and a) \protect\( \tilde{\alpha} =0.1\protect \),
b) \protect\( \tilde{\alpha} =0.4\protect \), c) \protect\( \tilde{\alpha} =0.7\protect \) and
d) \protect\( \tilde{\alpha} =1.5\protect \). The solid circles correspond 
to the instabilities at finite wavenumber due to the Hopf modes. The 
other lines are as in Fig. \ref{fig.gam}. The diamond in d) corresponds to the simulation 
in Fig. \ref{fig.num}.}

\label{fig.eta}
\end{figure}

In Figs. \ref{fig.gam} and \ref{fig.eta} we present the stability 
limits obtained for several values of \( \tilde{\nu} \) and \( 
\tilde{\alpha} \).  The short-dashed and the dotted lines correspond to the Hopf and 
saddle-node bifurcations, respectively, while the dot-dashed line is the curve above 
which the rolls become stable with respect to the hexagons.  We do not 
address their side-band instabilities.  The circles correspond to 
the results of the general stability analysis while the solid and 
dashed lines are the stability limits in the long-wave approximation, 
as given by (\ref{eq.disp}) (with eigenvalues either real or a complex 
conjugate pair).  The solid circles in Fig.  \ref{fig.eta} correspond 
to instabilities at finite wavenumber due to the Hopf modes.

Fig. \ref{fig.gam} shows the stability limits for $\tilde{\alpha}=0$ 
but $\tilde{\nu} \ne 0$, i.e.  the chiral symmetry is only broken at 
cubic order.  While for small values of the control parameter the 
long-wave analysis gives the correct stability limits, for larger 
$\mu$ a steady short-wave instability preempts the long-wave 
instability before it becomes oscillatory.  In this case the 
eigenvalues are complex for \( Q\rightarrow 0 \), but they split into 
two real eigenvalues for larger $Q$, one of which becomes positive 
(Fig. \ref{fig.disp}a).  As the coefficient \( \tilde{\nu} \) is 
increased the region in which the long-wave instability is relevant
decreases (Fig. \ref{fig.gam}b) and shrinks to almost zero  
(Fig. \ref{fig.gam}c).

When the quadratic gradient terms are different from zero the 
instability regions become asymmetric with respect to \( K =0\).  The 
system is, however, invariant under the change \( \tilde{\alpha} 
\rightarrow -\tilde{\alpha} \), \( K\rightarrow -K \), \(\alpha_i 
\rightarrow -\alpha_i\), and we will therefore consider only positive 
values of $\tilde{\alpha}$.  For small \( \tilde{\alpha} \) the 
region of the steady long-wave instability becomes smaller at one side 
of the bandcenter and larger at the other (Fig.  \ref{fig.eta}a).  As 
\( \tilde{\alpha} \) is increased, this region shrinks to zero for 
$K<0$ and disappears above the Hopf curve for $K>0$.  At this point 
all the instabilities are oscillatory (Fig. \ref{fig.eta}c).  For $\tilde{\alpha}=0.4$ and $\tilde{\alpha}=0.7$
(Fig. \ref{fig.eta}b,c) the stability limit is entirely given by the 
longwave results for $K>0$, but for still larger values of \( 
\tilde{\alpha} \) it becomes shortwavelength (Fig.  \ref{fig.eta}d).  
For $K<0$ there is a large region in which the instability is short-wave 
and oscillatory.  Close to the Hopf bifurcation the instability switches from
the translation modes  to the Hopf mode (cf. Fig. \ref{fig.disp}).
As $\tilde{\alpha}$ is increased the region in  which the instability is due to
the Hopf modes grows.

\section{Numerical simulations}

In order to study the nonlinear behavior arising from the 
instabilities, we have performed numerical simulations of Eqs.  
(\ref{eq.ampgen}).  A Runge-Kutta method with an integrating factor 
that computes the linear derivative terms exactly has been used.  
Derivatives were computed in Fourier space, using a two-dimensional 
fast Fourier transform (FFT).  The numerical simulations were done in 
a rectangular box of aspect ratio $2/\sqrt{3}$ with periodic boundary 
conditions.  This aspect ratio was used to allow for regular hexagonal 
patterns.

We start with a perfect hexagonal pattern with a wavenumber in the 
unstable region and add noise.  In all the cases we have considered 
the numerical simulations reproduce correctly the linear stability 
limits.  Over most of the parameter regime the nonlinear evolution of 
the instabilities is qualitatively very similar.  The perturbation 
growths (with or without oscillations, depending on the kind of 
instability) until it destroys the original hexagon pattern and then 
settles down to a stable periodic pattern.  Therefore all the 
instabilities appear to be subcritical.  Furthermore, it seems to be 
irrelevant whether the instability comes from the translation or the 
Hopf modes.  The branch switching does not change the value of the 
unstable wavenumber nor the frequency (cf. Fig. \ref{fig.disp}b,c) and 
the behavior of the dispersion relation 
at lower values of the perturbation wavenumber does not play a role. 
 For values of the control 
parameter for which rolls are unstable, the instabilities lead to a 
rotation of the original hexagonal pattern and to a change in its 
wavelength.  Usually penta-hepta defects appear in the process.  In 
the presence of rotation they annihilate each other quite fast, 
yielding a perfect pattern as the final state.  For larger control 
parameters rolls become stable and the side-band instabilities of the 
hexagons eventually lead to roll patterns, independently of the 
specific type of the instability. 

\begin{figure}

\centerline{
\epsfxsize=3.8cm\epsfbox{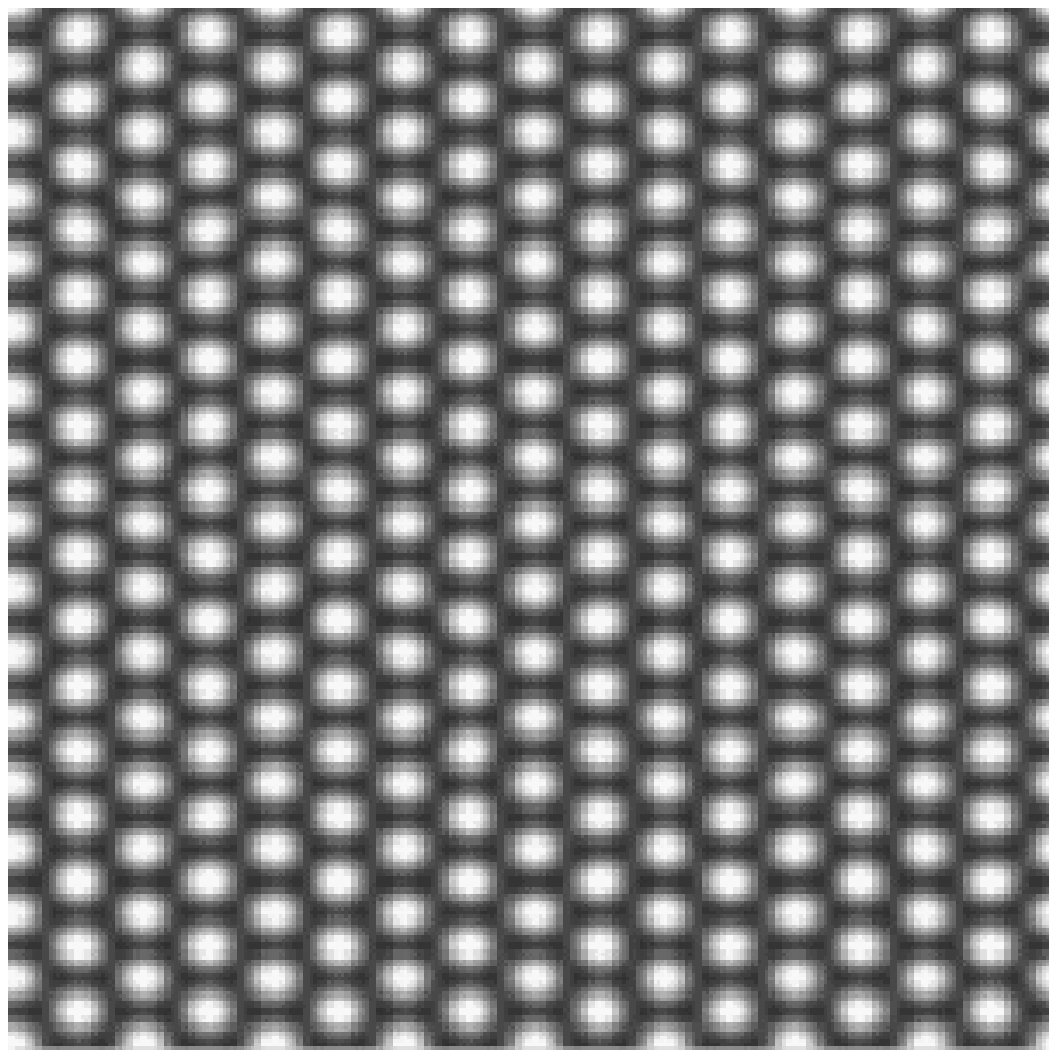}\hspace{.2cm}\epsfxsize=3.8cm\epsfbox{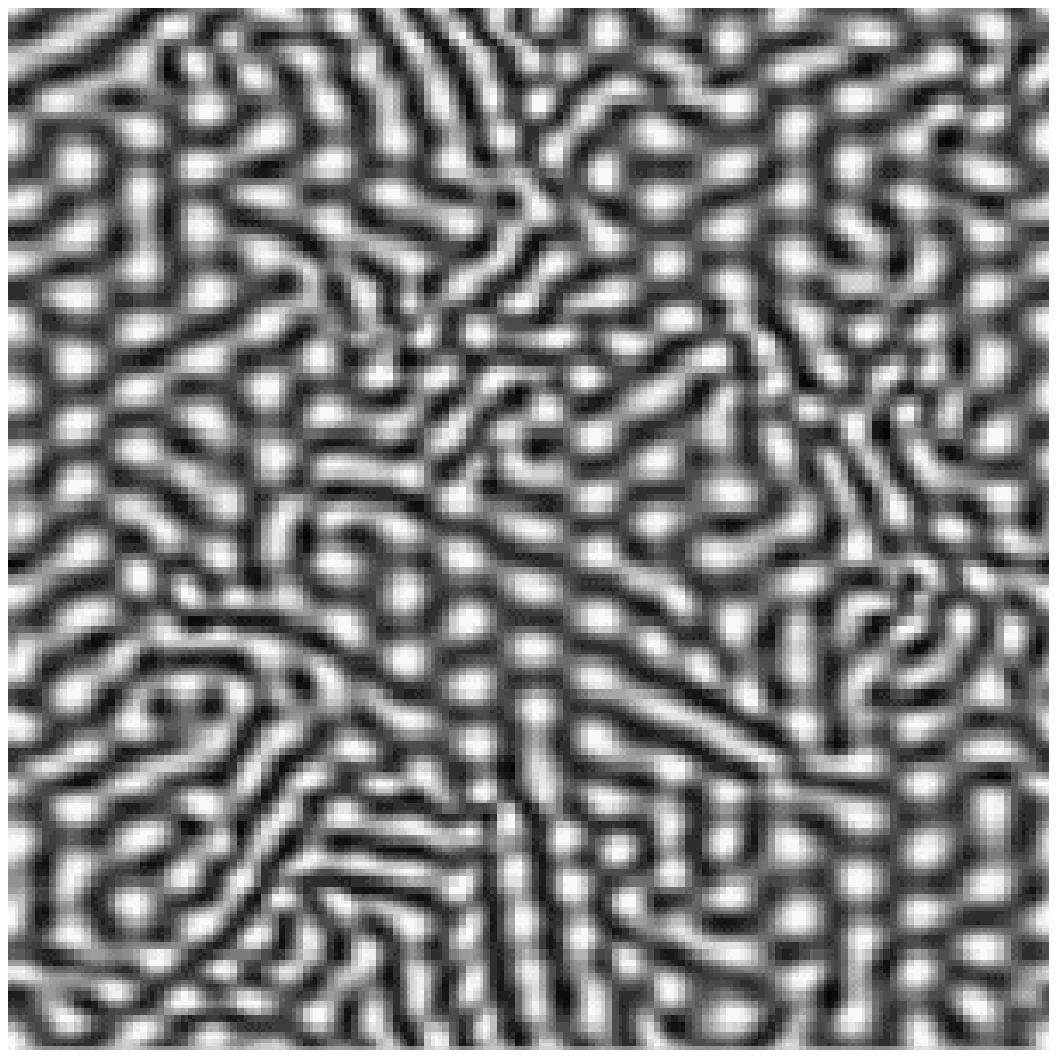}
\hspace{.2cm}\epsfxsize=3.8cm\epsfbox{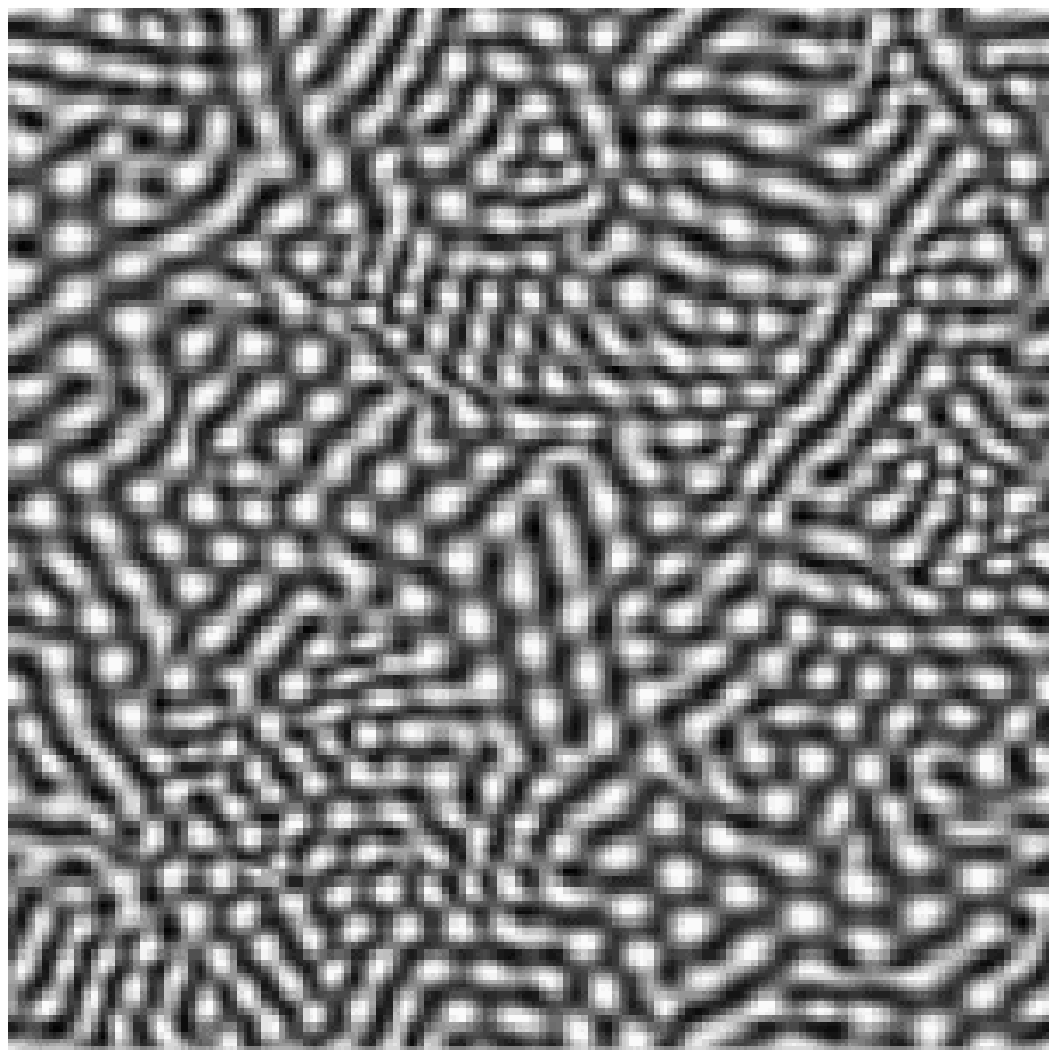}
}

\centerline{
\epsfxsize=3.8cm\epsfbox{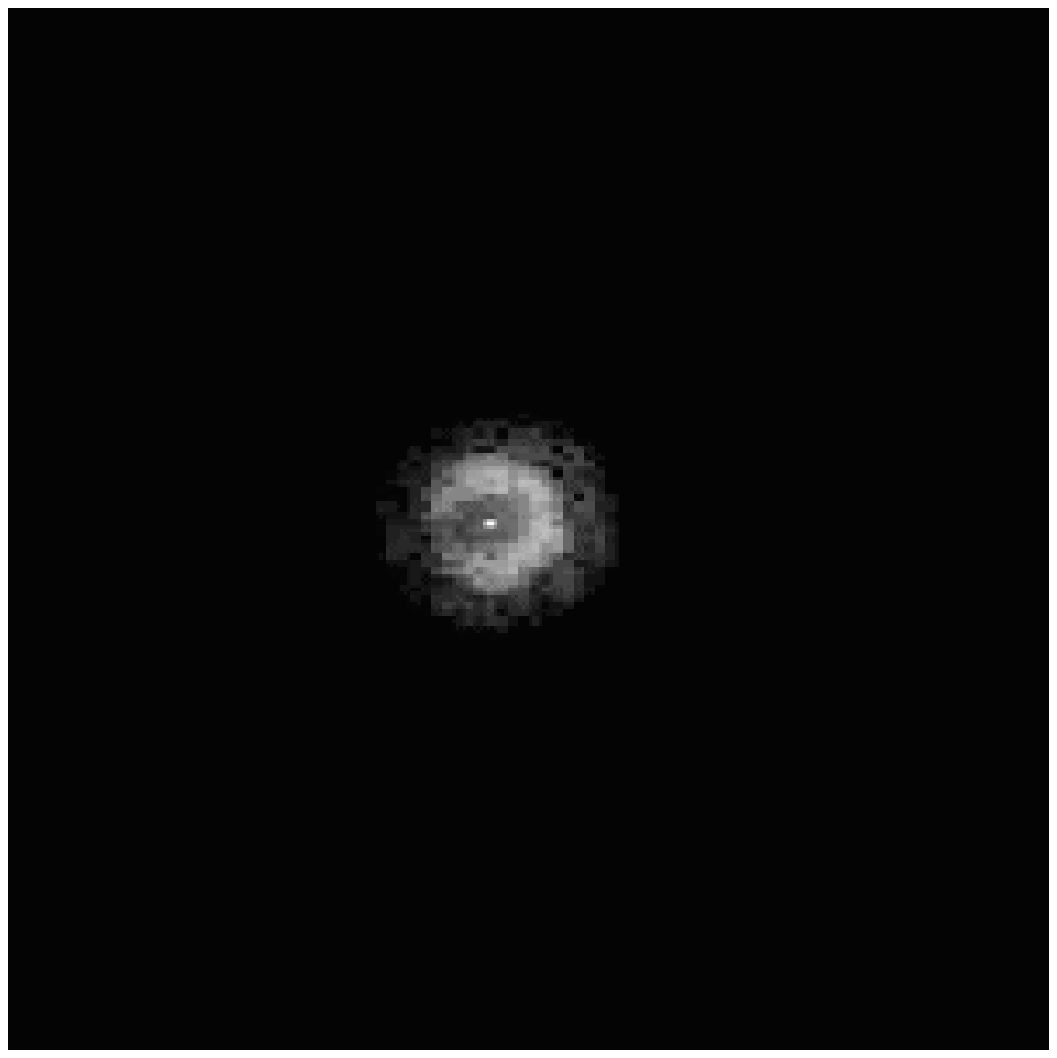}\hspace{.2cm}\epsfxsize=3.8cm\epsfbox{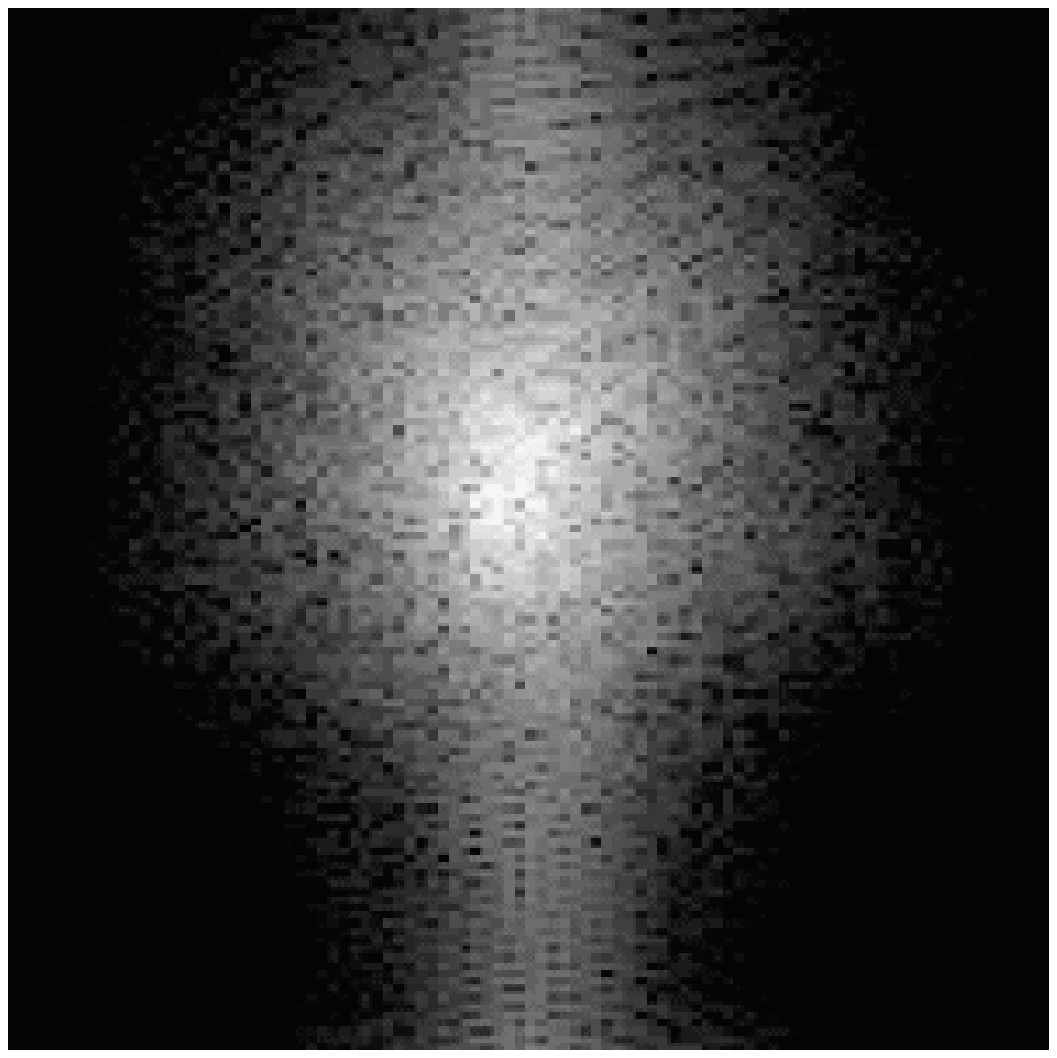}
\hspace{.2cm}\epsfxsize=3.8cm\epsfbox{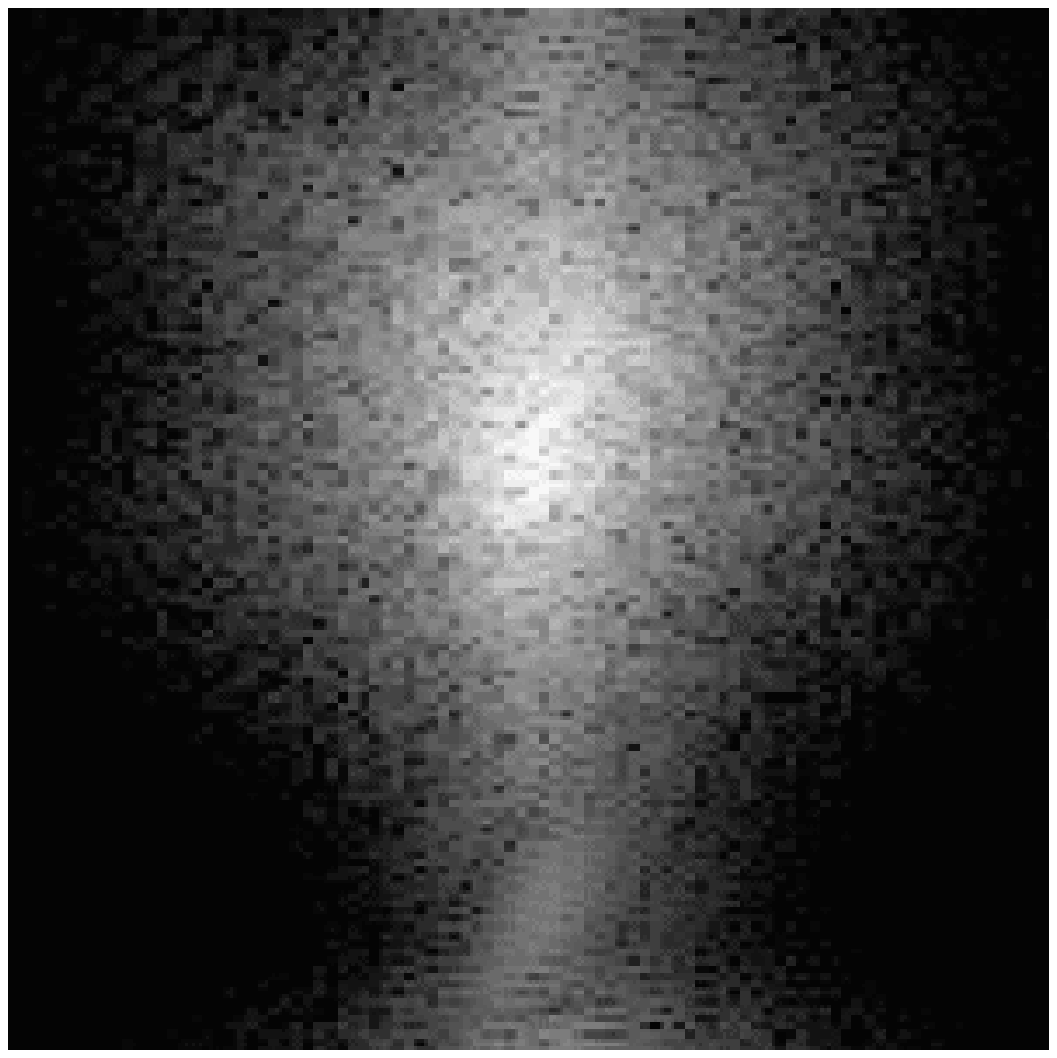}
}

\caption{Numerical simulations of Eqs. (\ref{eq.ampgen}) 
in a box of length $L=35$ ($K_{min}=0.1795$) with $128\times 128$ Fourier modes, 
for the values of the parameters $\nu=2$, $\tilde{\nu}=0.1$, $\alpha_1=\alpha_2=0$, 
$\tilde{\alpha}=1.5$, $\mu=3$ and $K=-4K_{min}$ (cf. 
Fig. \protect{\ref{fig.eta}d}). The top panels show a reconstruction of 
the hexagonal pattern $\Psi=\sum^{3}_{i=1} A_i e^{i{\bf k}^c_i \cdot {\bf 
x}}$ with $k_c=20K_{min}$ for the times $t=32$, $t=52$ and $t=56$. In the bottom 
panels the corresponding Fourier spectra of 
the amplitude $A_1$ are shown. Note the shifting of the peak of the distribution to ever larger $y$-components 
of the wavevector.}
\label{fig.num}
\end{figure}

\begin{figure}

\centerline{
\epsfxsize=7cm\epsfbox{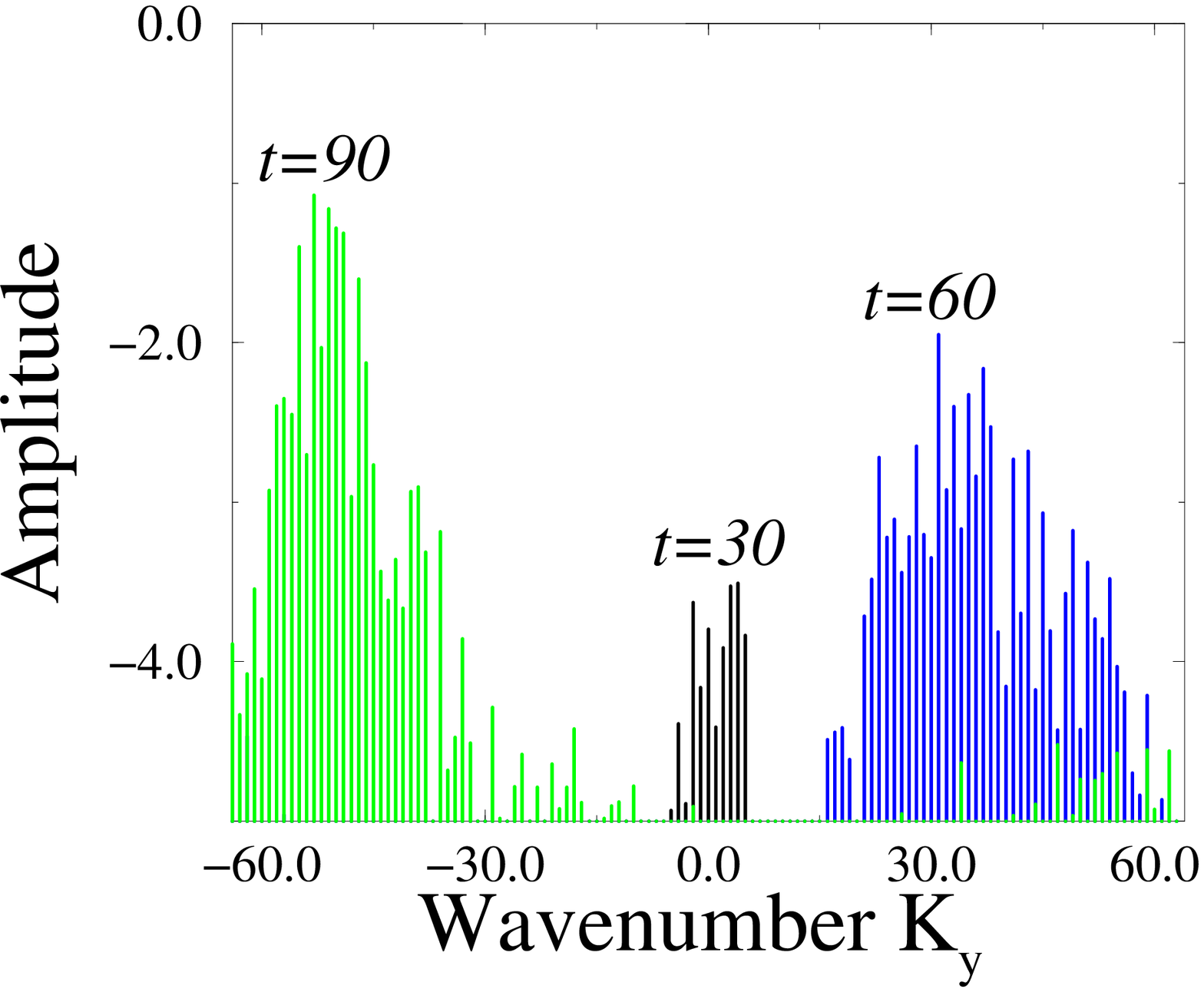}
}
\caption{A cut of the Fourier spectrum (in logarithmic scale) 
of the amplitude $A_1$ at $K_x=0$ for the same parameters as in Fig. \ref{fig.num}. 
As time progresses the peak of the distribution evolves towards higher wavenumbers and eventually 
reemerges at the negative end of the spectrum.}
\label{fig.four}
\end{figure}

For certain parameter values, however, more complicated behavior is 
found. This is shown in Fig. \ref{fig.num}, where we represent a 
reconstruction of the hexagonal pattern, $\Psi=\sum^{3}_{i=1} A_i 
e^{i{\bf k}^c_i \cdot {\bf x}}$ (top panel), as well as the corresponding Fourier spectrum of the 
amplitude $A_1$, $\hat{A}_1 (K)$ (bottom panel).  In this case the instability develops close to the 
initial wavenumber of the unstable hexagonal pattern 
(Fig. \ref{fig.num}a).  As time 
progresses, however, modes with ever increasing $y$-component of the 
wavevector are excited.  Independent of the maximal wavevectors 
retained in the simulations ($15.3 < K_{max} < 46$ with system 
size in the range $17.5 \le L \le 52.5$)  eventually the wavevectors 
with the largest possible $y$-components are excited and the peak in 
the spectrum 
displayed in Fig. \ref{fig.num}b reaches the top border of the 
figure.  Then the peak reemerges at the bottom border again, i.e.  the 
wavevectors have very large negative $y$-components. This is shown 
more clearly in Fig.  
\ref{fig.four} where a cross-section of the Fourier spectrum in the 
$y$-direction for $K_{x}=0$ is shown for three times. At time $t=90$ 
most of the excited modes have already reemerged at (large) negative 
values of $K_{y}$. Obviously, in these simulations the solutions cease 
to be numerically resolved already well before $t=60$.  
Within the Ginzburg-Landau equations 
(\ref{eq.ampgen}) the curve of marginal modes corresponds to a 
vertical line in the Fourier spectrum in Fig. \ref{fig.num}. Thus, the 
excited modes lie predominantly along the critical curve and 
the numerically observed 
behavior suggests   that the correct evolution of the pattern 
would involve  a trend towards a rotation of the pattern  and
a spreading of the Fourier modes over the circle of 
marginal modes. Such 
dynamics reflect explicitly the isotropy of the system. They cannot be 
captured within the Ginzburg-Landau equations, which break the 
isotropy through the choice of the wavevectors corresponding to the 
amplitudes $A_i$.
To represent dynamics as suggested in Fig. \ref{fig.num} correctly, models that retain the 
isotropy have to be 
used.  This motivates the use of Swift-Hohenberg-type models 
\cite{SwHo77,XiGu94,MiPe92}\footnote{Formally, isotropy can be 
recovered by modifying the gradient terms in Eq. (\ref{eq.ampgen}) 
\cite{GuOu94,Gr96}. However, this requires that the amplitudes be allowed 
to vary rapidly in space.}.  Investigations of the complex dynamics
that can arise from instabilities identified here have been 
performed in \cite{SaRi99}. They show indeed a bistability between the ordered hexagons and a spatio-temporally chaotic state with an almost isotropic Fourier spectrum.

\section{Conclusion}

In this article we have analyzed the effect of chiral 
symmetry-breaking on the stability of hexagonal patterns.  Such 
patterns arise, for instance, in non-Boussinesq Rayleigh-B\'enard 
convection and in Marangoni convection, where the chiral symmetry can 
be broken by rotating the system.  Focussing on the regime near 
threshold we have used the appropriate Ginzburg-Landau equations for 
the three modes making up the hexagon pattern.  The chiral symmetry 
breaking introduces an asymmetry between the cubic coupling coefficients 
as well as a new nonlinear gradient term.  The general linear 
stability analysis of these equations revealed long-wave  
as well as short-wave instabilities.  The long-wave instabilities, 
which are captured with coupled phase equations, can be steady or 
oscillatory. For all parameter regimes investigated, the short-wave 
instabilities arise for larger values of the control parameter, but 
below the transition to oscillating hexagons. 
They can be due to the translation or the Hopf modes. 
In the latter case they are always oscillatory.

In contrast to the K\"uppers-Lortz instability of stripe patterns 
\cite{KuLo69}, no regime was identified in which hexagon patterns 
become unstable at all wavelengths.  Nevertheless, persistent 
irregular dynamics of disordered hexagon patterns can apparently arise 
from the short-wave instability.  Our numerical simulations of the 
Ginzburg-Landau equations indicate that the nonlinear evolution 
ensuing from the instability tends to introduce modes with wavevectors 
covering the whole critical circle.  Of course, such a state in which 
the Fourier modes are distributed almost isotropically over the 
critical circle cannot be described by Ginzburg-Landau equations, 
since they break the isotropy at the very outset.  This suggests the 
use of Swift-Hohenberg-type equations, which preserve the isotropy of 
the system.  They are often used as truncated model equations 
(e.g. \cite{BeFr99}) to study the qualitative behavior of systems, but can 
under certain conditions also be derived from the basic (fluid) 
equations as a long-wave description \cite{Kn90,Co98,MaSaunpub}.  Recently, 
in such investigations of hexagons with broken chiral symmetry 
spatio-temporally chaotic states have been found to arise from the 
corresponding oscillatory short-wave instability \cite{SaRi99}.  As in 
our simulations of the Ginzburg-Landau equations the spatio-temporal 
chaos persists although for the same parameters there exist also 
stable ordered hexagon patterns.  This bistability is somewhat 
reminiscence of the coexistence of spiral-defect chaos and ordered 
roll convection in Rayleigh-B\'enard convection without rotation 
\cite{MoBo93}.  In the Swift-Hohenberg model the oscillatory short-wave 
instability can also lead to a supercritical bifurcation to hexagons 
that are modulated periodically in space and time \cite{SaRi99}.  No such 
state could be identified in the Ginzburg-Landau equations discussed 
here.

Our results suggest that rotation may induce irregular dynamics in 
hexagonal convection patterns quite close to threshold. So far, 
disordered hexagon patterns (without broken chiral symmetry) 
have been found in Marangoni convection \cite{ThOr95} far from threshold and also
in experiments on chemical Turing patterns \cite{OuSw91}. In the
latter case they appear to be due to the competition with the stripe 
pattern in a bistable regime.

From previous work it is well known that the chiral symmetry breaking 
delays the transition from hexagons to stripe patterns. More 
specifically, the steady bifurcation to the unstable mixed state is 
replaced by a Hopf bifurcation to a state of coherently oscillating 
hexagons \cite{Sw84,So85,MiPe92}. Their side-band instabilities can  
be investigated with the same Ginzburg-Landau equations as discussed 
here \cite{EcRiunpub}.  

\acknowledgments

We gratefully acknowledge interesting discussions with F. Sain, M. 
Silber and C. P\'erez-Garc\'{\i}a. The numerical simulations were performed with a modification of a code by G.D. Granzow. This work was supported by D.O.E. Grant DE-FG02-G2ER14303 and NASA Grant NAG3-2113. 

%\bibliography{/home2/hermann/.bibfiles/journal}
%\bibliography{journal}

\end{document}